\documentclass[review]{elsarticle}
\usepackage{lineno,hyperref}
\usepackage{color}
\usepackage{amsmath,amsfonts,amssymb,esvect,ulem}
\usepackage{float}
\usepackage{graphicx}
\usepackage{epsfig}

\usepackage{epstopdf}
\usepackage{subcaption}
\modulolinenumbers[5]

\journal{Journal of Computational Physics}

\newcommand{\st}{\sigma_\mathrm{t}}

\newcommand{\pn}{P$_N$}
\newcommand{\dn}{D$_N$}
\newcommand{\tp}[1]{TP$_{#1}$}
\newcommand{\ppz}{\partial_x}

\newcommand{\psii}[1]{\phi_\ensuremath{{#1}}}

\newcommand{\sn}{S$_N$}

\newcommand{\TAMU}{Texas A\&M University}

\newcommand{\sigmat}{\sigma_\mathrm{t}}
\newcommand{\sigmas}{\sigma_\mathrm{s}}

\begin{document}

\begin{frontmatter}

\title{Moment Closures Based on Minimizing the Residual of the \pn~Angular Expansion in Radiation Transport}

\author[mymainaddress]{Weixiong Zheng}
\ead{zwxne2010@tamu.edu}

\author[mymainaddress]{Ryan G. McClarren\corref{mycorrespondingauthor}}
\cortext[mycorrespondingauthor]{Corresponding author}
\ead{rgm@tamu.edu}

\address[mymainaddress]{Nuclear Engineering, \TAMU,~College Station, TX 77843-3133}

\begin{abstract}
	In this work we present  two new closures for the spherical harmonics (\pn) method in slab geometry transport problems. Our approach begins with an analysis of the squared-residual of the transport equation where we show that the standard truncation and diffusive closures do not minimize the residual of the \pn~expansion.  Based on this analysis we derive two models, a moment-limited diffusive (ML\dn) closure and a transient \pn~(\tp{N}) closure that attempt to address shortcomings of common closures.  The form of these closures is similar to flux-limiters for diffusion with the addition of a time-derivative in the definition of the closure.  Numerical results on a pulsed plane source problem, the Gordian knot of slab-geometry transport problems, indicate that our new closure outperforms existing linear closures.  Additionally, on a deep penetration problem we demonstrate that the \tp{N}~closure does not suffer from the artificial shocks that can arise in the M$_N$ entropy-based closure. Finally, results for Reed's problem demonstrate that the \tp{N}~solution is as accurate as the P$_{N+3}$ solution. {We further extend the T\pn\ closure to 2D Cartesian geometry. The line source test problem demonstrates the model effectively damps oscillations and negative densities.}

\end{abstract}


\end{frontmatter}

\linenumbers

\section{Introduction}
The Boltzmann equation is used to describe particle transport in several applications, e.g.\ neutron transport \cite{glasstone}, thermal radiative transfer \cite{pomraning1973equations}, rarefied gas dynamics \cite{Grad:1949wi}, and charged-particle transport in semiconductors \cite{markowich2012semiconductor}, to name a few. Solving the transport equation is challenging due to the seven-dimensional phase space. In this work we deal with a simple transport model with a linear collision operator, though our methods could be extended to more complicated transport processes. 

For the angular discretization a commonly used method is the discrete ordinates method (\sn),\ which solves the transport equation for several selected discrete angles. However, there are physical situations where \sn\ encounters difficulties. One example is in multi-dimensional applications, when the medium weakly interacts with the particles. In this case the solution along each ordinate are not coupled, which leads to the well-known ``ray effects" in the solution \cite{Mathews:1999uv}. Ray effects can remain present even when the number of discrete angles is large \cite{Morel:2003vt}. 

Another approach is to use a truncated spherical harmonics expansion (\pn)\ on the angular variable. The \pn~method is a spectral method in angle based on a linear expansion of angular flux, yielding a hyperbolic system of partial differential equations (PDEs) for the expansion coefficients, or equivalently, the moments. For smooth angular dependence, the method has spectral convergence. Also, the spherical harmonics expansion are rotationally invariant, in contrast to \sn,\ thereby avoiding ray effects.

Nevertheless, in time-dependent problems, truncating the basis at a finite order $N$\ and assuming moments with higher orders to be zero\footnote{The resulting closure is referred to as zero closure in this paper.} causes the solution  to approximate the radiation as a finite number of waves moving with different speeds determined by the eigenvalues of the coefficient matrix of spatial derivative terms. These wave speeds are less than the real speed of the particles \cite{brunner_app_rad_trans,mccfpn09,McClarren:2008cq}.\ The artifacts in the solution that arise from the discrete wave speeds are referred to as ``wave effects". Wave effects can induce oscillations and negativity densities in the solution \cite{mccfpn09,coryentropy}. Though one could increase the order of the  \pn\ approximation to mitigate the oscillations, for a finite order of approximations, there are always certain physical situations where negative particle densities can occur \cite{McClarren:2008hu,mccdissertation}.

Here we briefly summarize the recent efforts to improve the \pn~method.\ A linear diffusive closure was developed independently by Schafer, Frank and Levermore \cite{levermoredn}\ for general orders of expansion and by Kyeong and Holloway specifically for P$_3$ \cite{p3qs}. Solutions using this closure demonstrated faster convergence to the transport solution. Additionally, Olson introduced a modification of the coefficients for highest moment equations as introduced in the P$_{1/3}$\ equation so the maximum speed of the waves is fixed to be the correct particle speed \cite{Olson:2000vq}. 

Other work has investigated closures based on the solution of an optimization problem based on an entropy (called the M$_N$ method) \cite{cory_hauck_closures,brunnerentropy} or the positivity of the particle density (called the positive \pn~method) \cite{ppn}. These methods do have some benefits, such as guaranteed positive solutions. The numerical solution of the resulting equations using high-order expansions is computationally prohibitive because an optimization problem must be solved for each spatial degree of freedom in each time step. Additionally, the optimization problem resulting from high-order M$_N$ expansions is ill-conditioned. Furthermore, in some test problems, entropy-based closures can cause artificial  shocks to develop in the solution  \cite{cory_hauck_closures,brunnerentropy}. 

Inspired by the usage of artificial viscosity in hydrodynamics, Hauck and McClarren introduced a filtering process, which is interpreted as artificial viscosity in angle \cite{mccfpn09,McClarren:2010de}.\ The basic idea is to introduced a new spherical harmonics basis which introduces dampening on high order moment coefficients. Though it outperforms conventional closures, it cannot guarantee positivity of the particle density because it is a linear closure. Radice et al\ introduced a new form of the filter that was independent of the mesh and time step \cite{fpn_radice} and applied it to radiative transfer problems in astrophysics. {On the other hand, it is also found that the filtering introduced by McClarren and Hauck does not smoothly transition to zero when increasing the \pn\ angular order. Ahrens et al\ introduce a cubic filter to address the problem\cite{ahrens_fpn}.}


In this work, we introduce two new closures that are designed to improve the residual of \pn~expansions in 1D slab geometry. We start off defining a functional based on the angularly-integrated squared residual. By examining the minimizer of the functional, we arrive at two nonlinear closures based on the residual. Both approaches write the closure as a rational function times the derivative of a moment, similar to flux limited diffusion. We analytically demonstrate one closure will bound the magnitude of highest moment by the scalar flux. The other closure is a modification that involves the zeroth moment of the angular flux. Further, we numerically demonstrate the high accuracy of the closures in a problem with strong wavefronts. On Reed's problem \cite{reed_1971} we show that our new closure converges to the transport solution faster than diffusive closures or the standard truncation.

\section{Derivation of the Method}
\subsection{Error functional derivation of the P$_N$ equations}\label{s:derive}
We begin with an energy-independent transport equation for neutral particles in slab geometry given by \cite{glasstone} 
\begin{equation}\label{te}
\frac{1}{v}\frac{\partial\psi(x,\mu,t)}{\partial t}+\mu \frac{\partial}{\partial x} \psi(x,\mu,t)+\sigma_\mathrm{t}\psi(x,\mu,t)=q(x,\mu,t).
\end{equation}
In this equation the angular flux of particles is given by $\psi(\mathbf{r},\mathbf{\Omega},t)$ with units of particles per area per time. Our notation is standard with $x \in \mathbb{R}$ being the spatial variable, $\mu \in [-1,1]$ as the cosine of the angle between the slab normal and the direction of flight, and $t$ as the time variable. The macroscopic total interaction cross-section with units of inverse length is given by $\sigmat$, and $q$ contains the prescribed source, $Q(x,\mu)$, and the scattering source:
\begin{equation} \label{eq:scat_source}
q = \frac{Q}{2} + \frac{\sigmas}{2}\phi,
\end{equation}
with the scalar flux, $\phi(x,t)$, defined as
\begin{equation} \label{eq:scalarflux}
\phi(x, t)  = \int\limits_{-1}^{1} d\mu\, \psi(x,\mu,t).
\end{equation}

In order to solve Eq.~(\ref{te}) one needs to apply discretizations in space, angle, and time.  In this work we focus on the angular discretization, in particular we will expand the angular dependence in Legendre polynomials as 
\begin{equation} \label{eq:Pn_expansion}
\psi(x,\mu,t) = \sum_{l=0}^\infty C_l \phi_l(x,t) P_l(\mu),
\end{equation}
where the Legendre polynomials are given by
\begin{equation} \label{eq:legPoly}
P_l(\mu) = {1 \over 2^l l!} {d^l \over d\mu^l } \left[ (\mu^2 -1)^l \right].
\end{equation}
Here, 
$\phi_l(x,t)$ is an expansion function, and $C_l$ is a normalization constant given by
\[ C_l = \left(\int_{-1}^{1} d\mu\,P_l(\mu) P_l(\mu)\right)^{-1}.\]
This technique is known as the $P_n$ method, and generalizes to general three-dimensional geometries by making the expansion functions spherical harmonics \cite{glasstone}.

The typical way that the expansion functions and normalization constants are generated is via a Galerkin procedure where one assumes a Legendre expansion of the angular flux, plugs it into the transport equation, and integrates the result against different Legendre polynomials. An alternative derivation involves defining a {error functional} measuring the difference of angular flux, $\psi$~and the spherical-harmonics-reconstructed angular flux using an expansion that is truncated beyond the  $l=N$ moment, $\bar{\psi}_N$ \cite{mccfpn09}. We define the errror functional as the integrated square of the difference between the true angular flux and the truncated expansion:
\begin{equation}\label{cstf1}
J_1(\mathbf{\Omega})=\int\limits_{4\pi}d\Omega~(\psi-\bar{\psi}_N)^2,
\end{equation}
where
\begin{equation}\label{cstf}
\bar{\psi}_N(\mathbf{\Omega})=\sum_{l=0}^N C_l \phi_l(x,t) P_l(\mu).
\end{equation}
In order to minimize the functional in Eq.~\eqref{cstf},~one forces $\partial J_1/\partial\phi_l=0$, leading to
the expansion coefficients being given by
\begin{equation}
\phi_l=\int\limits_{-1}^{1}d\mu\,\psi(x,\mu,t)P_l(\mu).
\end{equation}

Using this definition for the expansion coefficients, we take a certain Legendre polynomial and integrate it with the transport equation, Eq.~(\ref{te}), over $\mu$ to get:
\begin{equation}
\frac{1}{v}\frac{\partial}{\partial t} \phi_l+\frac{l}{2l+1}\frac{\partial\phi_{l-1}}{\partial x} +\frac{l+1}{2l+1}\frac{\partial\phi_{l+1}}{\partial x}+ (\sigma_{t} - \sigmas\delta_{l,0} )\phi_l=q_\mathrm{ext}\delta_{0,l},~l=0,1,\cdots N
\end{equation}
where $\displaystyle q_\mathrm{ext}(x)=\int\limits_{-1}^1d\mu~Q(x,\mu)$. 
This system is not closed in the sense that the equation for the $N$th moment includes the $N+1$ moment, which is not included in our truncated expansion. A common closure is to set $\psii{N+1}=0$. Thereafter, the closed \pn~equation system can be described as:
\begin{multline}\label{pne}
\frac{1}{v}\frac{\partial}{\partial t}\phi_l+\frac{l}{2l+1}\frac{\partial\phi_{l-1}}{\partial x}+\frac{l+1}{2l+1}\frac{\partial\phi_{l+1}}{\partial x}(1-\delta_{N,l})\\+(\sigma_\mathrm{t}-\sigma_\mathrm{s}\delta_{0,l})\phi_l
=q_\mathrm{ext}\delta_{0,l},~l=0,1,\cdots,N.
\end{multline}
These equations are the standard \pn\, equations.  We will now change the derivation to use a functional that minimizes the residual in the transport equation given a particular expansion.


\subsection{A functional based on the squared-residual}
%
Rather than basing the functional on the squared difference between the expansion and the true solution, one could also measure the squared residual of the \pn~approximation as:
\begin{equation}
J(\{\psii{l'}:l'=0,\cdots\})=\int\limits_{-1}^{1}d\mu~R^2,
\end{equation}
where $R$~is the residual computed when the expanded flux to order $N$ is plugged into the transport equation with an isotropic source. For  simplicity, we consider the pure absorber problem (though removing this restriction leads to the same results) leading to the definition of residual as
\begin{align}\label{res}
R\equiv\mathcal{L}\bar{\psi}_N(\mu)-q(\mu)
=\left(\frac{1}{v}\partial_t+\mu\ppz+\st\right)\sum\limits_{l'=0}^{N}\frac{2l'+1}{2}\psii{l'}P_{l'}(\mu)-\frac{q_\mathrm{ext}}{2},
\end{align}
where the transport operator $\mathcal{L}$~is defined as:
\begin{equation}
\mathcal{L}\equiv\frac{1}{v}\partial_t+\mu\ppz+\st.
\end{equation}

In order to minimize the functional $J$, we focus on finding moment sets which make $\partial J/\partial\psii{l}=0$ for all $l$. Through this path, we could gain an insight into the impact on the residual due to the closure.

Taking the functional derivative of Eq.~\eqref{res}~leads to:
\begin{multline}\label{func}
\frac{\partial J}{\partial\psii{l}}=(2l+1)\st\int\limits_{-1}^{1}d\mu~RP_l(\mu)~+ 
(2l+1)\frac{\partial}{\partial\psii{l}}\left[\frac{\partial}{\partial x}\psii{l}\right]
\int\limits_{-1}^{1}d\mu~R\mu P_l(\mu)+\\(2l+1)\frac{1}{v}\frac{\partial}{\partial\psii{l}}\left[\frac{\partial}{\partial t}\psii{l}\right]\int\limits_{-1}^{1}d\mu~RP_l(\mu).
\end{multline}

Note that for $l\leq N$, the following identity holds
\begin{equation}\label{pnn}
\int\limits_{-1}^1d\mu~RP_l(\mu)=\frac{1}{v}\partial_t\phi_l+\frac{l}{2l+1}\frac{\partial\phi_{l-1}}{\partial x}(1-\delta_{0,l})+\sigma_{t}\phi_l+\frac{l+1}{2l+1}\frac{\partial\phi_{l+1}}{\partial x}-q_\mathrm{ext}\delta_{0,l}.
\end{equation}
Comparing Eq.~\eqref{pnn}~with Eq.~\eqref{pne},~one sees that the integral is equal to zero, i.e.
\begin{equation}\label{pnn2}
\int\limits_{-1}^1d\mu~RP_l(\mu)=0,\qquad l\leq N.
\end{equation}

Also, by using recurrence relation of Legendre polynomial, one has:
\begin{equation}\label{recurr}
\int\limits_{-1}^{1}d\mu~R\mu P_l(\mu)=\frac{l}{2l+1}\int\limits_{-1}^{1}d\mu~\mu P_{l-1}(\mu)R+\frac{l+1}{2l+1}\int\limits_{-1}^{1}d\mu~\mu P_{l+1}(\mu)R.
\end{equation}

Therefore, Eq.~\eqref{func}~can be rewritten as:
\begin{multline}\label{func2}
\frac{\partial J}{\partial\psii{l}}=(2l+1)\left(\st+\frac{1}{v}\frac{\partial}{\partial\psii{l}}\left[\frac{\partial}{\partial t}\psii{l}\right]\right)\int\limits_{-1}^{1}d\mu~RP_l(\mu)+\\
\frac{\partial}{\partial\psii{l}}\left[\frac{\partial}{\partial x}\psii{l}\right]
\left(l\int\limits_{-1}^{1}d\mu~RP_{l-1}(\mu)+(l+1)\int\limits_{-1}^{1}d\mu~RP_{l+1}(\mu)\right).
\end{multline}

When $l < N$,~
plugging Eq.~\eqref{pnn2}~back into Eq.~\eqref{func2} leads to:
\begin{equation}
\frac{\partial J}{\partial\psii{l}}=0, \qquad l < N.
\end{equation}
That is,  all of the P$_N$ equations minimize the squared-residual for $\phi_l$ for all $l<N$. This is why the omission of the scattering term does not affect our results: the scattering term only appears in the $l=0$ equation. It is in the $l=N$ equation where the closure enters. We will now explore what that equation tells us.

For $l=N$,~the same substitution, omitting the algebraic process, results in:
\begin{align}\label{func3}
\frac{\partial J}{\partial\psii{N}}=(N+1)
\frac{\partial}{\partial\psii{N}}\left[\frac{\partial}{\partial x}\psii{N}\right]
\left(\int\limits_{-1}^{1}d\mu~RP_{N+1}(\mu)\right).
\end{align}
Therefore, expanding the integral term in Eq.~\eqref{func3}~gives us the final expression for the $N^\mathrm{th}$~order functional derivative:
\begin{multline}\label{func4}
\frac{\partial J}{\partial\psii{N}}=(N+1)
\frac{\partial}{\partial\psii{N}}\left[\frac{\partial}{\partial x}\psii{N}\right]
\left(\frac{1}{v}\partial_t\phi_{N+1}+\frac{N+1}{2N+3}\frac{\partial\phi_{N}}{\partial x}\right.\\
\left.+\sigma_{t}\phi_{N+1}+\frac{N+2}{2N+3}\frac{\partial\phi_{N+2}}{\partial x}\right).
\end{multline}
This equation can tell us the impact of a closure on the residual: to this equation we can substitute in a closure and see how it effects the derivative of the squared residual.


\subsection{Discussion on two conventional closures}\label{s:above}

\subsubsection{Zero closure}
Introducing the zero closure ($\psii{\mathcal{M}}=0,~\mathcal{M}>N$) into Eq.~(\ref{func4}) gives the following:
\begin{equation}\label{bd}
\frac{\partial J}{\partial\psii{N}}=(N+1)
\frac{\partial}{\partial\psii{N}}\left[\frac{\partial}{\partial x}\psii{N}\right]
\frac{N+1}{2N+3}\frac{\partial\phi_{N}}{\partial x}
.
\end{equation}
This equation indicates that the squared-residual will be minimized only if the spatial derivative of $\phi_N$ is zero.  This restriction is not expected to be satisfied in general problems.


\subsubsection{Diffusive closure}
Levermore et al.~suggested a diffusive closure which takes a form similar to Fick's law for the relationship between $\psii{N+1}$~and $\psii{N}$\cite{levermoredn}:
\begin{equation}\label{fick}
\psii{N+1}=-\frac{1}{\st}\frac{N+1}{2N+3}\ppz\psii{N}.
\end{equation}
They, therein, name the corresponding system D$_N$~in the sense that the closure is essentially taking the definition of diffusion to a high-order closure. Also, Oh and Holloway independently derived a low order D$_2$~method for transient problem by assuming the closed moment $\psii{N+1}$~is time-independent such that one could directly gain the Fick's law-like relationship in Eq.~\eqref{fick}\cite{p3qs}.~They name the method P$_3$QS, short for P$_3$~quasi static, because of the approximation used to find the closure. 

Substituting Eq.~\eqref{fick}~into Eq.~\eqref{func4}, we get:
%
\begin{equation}\label{bd2}
\frac{\partial J}{\partial\psii{N}}=(N+1)
\frac{\partial}{\partial\psii{N}}\left[\frac{\partial}{\partial x}\psii{N}\right]
\left(-\frac{N+2}{2N+3}\frac{1}{v}\partial_t\left(\frac{1}{\st}\ppz\psii{N}\right)+\frac{N+2}{2N+3}\frac{\partial\phi_{N+2}}{\partial x}\right).
\end{equation}
This result indicates where the \dn\, closure might be accurate. It will minimize the squared-residual when the time derivative of  $\phi_N$ is zero and when the spatial derivative of the $\phi_{N+2}$ is zero.  We cannot know for a general problem what the derivative of $\phi_{N+2}$ will be. Nevertheless, we can predict when transients have died out in a particular problem.  In such an occasion we predict that the \dn\, closure will be superior to the zero closure because the derivative of the $\phi_{N+2}$ moment impacts the residual, rather than the $\phi_N$ moment in the zero closure.

\subsection{Two new closures}
\subsubsection{Approximations on higher moments}
Equation ~\eqref{func4} indicates that we should seek a closure such that:
\begin{equation}\label{func5}
\frac{1}{v}\partial_t\phi_{N+1}+\frac{N+1}{2N+3}\frac{\partial\phi_{N}}{\partial x}
+\sigma_{t}\phi_{N+1}+\frac{N+2}{2N+3}\frac{\partial\phi_{N+2}}{\partial x}=0,
\end{equation}
which is equivalent to introducing a higher order \pn~approximation without changing the truncation order. The closure, leading to zero functional derivative in moment space, would potentially lead to a minimized residual of the \pn~approximation. However, this is not feasible practically since truncating at a certain order $N$~would lead to the loss of information of higher orders, e.g.~$\psii{N+2}$. The value of Eq.~\eqref{func5}~is that it indicates how one could close the system to minimize the residual in moment space.

Formally, we can rewrite Eq.~\eqref{func5} to implicitly define a closure as
\begin{equation}\label{oricl}
\psii{N+1}=-\frac{1}{\sigma_\mathrm{t}+\displaystyle\frac{\partial_t\psii{N+1}}{v\psii{N+1}}+\frac{(N+2)}{(2N+3)\psii{N+1}}\ppz\psii{N+2}}\frac{N+1}{2N+3}\ppz\psii{N}.
\end{equation}

\subsubsection{A moment-limited closure}
The closure indicated by Eq.~(\ref{oricl}) shares a similar form with diffusive closure, except, there are additional terms added to correct the closure. Though it is still a formal closure since it depends on the value of $\psii{N+1}$\ and $\psii{N+2}$, it implies adding spatial and temporal flux limiters to the diffusive closure could help minimize $\partial J/\partial\psii{N}$. Therefore, we propose the following closure:
\begin{equation}\label{ml}
\psii{N+1}=-\frac{1}{\sigma_\mathrm{t}+\displaystyle\left|\frac{\partial_t\psii{0}}{v\psii{0}}\right|+\left|\frac{\alpha\partial_x\psii{N}}{\psii{0}}\right|}\frac{N+1}{2N+3}\ppz\psii{N}.
\end{equation}

%
A desirable feature is that if a proper $\alpha$ is used, one could prove that this form limits the magnitudes of the closure as follows:
\begin{align}
|\phi_{N+1}|=\frac{1}{\sigma_\mathrm{t}+\displaystyle\left|\frac{\partial_t\psii{0}}{v\psii{0}}\right|+\alpha\left|\frac{\partial_x\psii{N}}{\psii{0}}\right|}\frac{N+1}{2N+3}\left|\ppz\psii{N}\right|\\\nonumber
\leq \frac{1}{\displaystyle\alpha\left|\frac{\partial_x\psii{N}}{\psii{0}}\right|}\frac{N+1}{2N+3}\left|\ppz\psii{N}\right|=\frac{N+1}{\alpha(2N+3)}|\phi_0|
\end{align}

For instance, fixing $\alpha$~at $(N+1)/(2N+3)$~would result in:
\begin{equation}
|\psii{N+1}|<|\psii{0}|
\end{equation}

That is similar to the situation of limiting current to the scalar flux to stabilize the system in moment space. We, therefore, name this approach the moment-limited diffusive (MLD) closure. 

\subsubsection{A modification: transient \pn\ closure}
The moment limited closure could be modified to use $\psii{0}$, instead of $\psii{N}$ in the closure. Specifically, the modified closure is expressed as:
\begin{equation}\label{closure3}
\psii{N+1}=-\frac{1}{\sigma_\mathrm{t}+\displaystyle\left|\frac{\partial_t\psii{0}}{v\psii{0}}\right|+\left|\frac{\alpha\partial_x\psii{0}}{\psii{0}}\right|}\frac{N+1}{2N+3}\ppz\psii{N}
\end{equation}
There are two motivations for this choice. On one hand, this selection is to make a form similar to  
a high order extension of flux-limited diffusion with an additional constraint on the temporal evolution of the solution (i.e. the $\partial_t\psii{0}$~term). Moreover, in multidimensional problems $\psii{0}$ is the only moment that is a scalar, making the extension to full spherical harmonics closures straightforward. {In contrast, extending ML\dn\ to multi-D requires individual estimates of the spatial limiters for each single $(N+1)^\mathrm{th}$\ moment equation. For instance, for MLD$_3$\ in 2D with moments generated from complex-value spherical harmonics, four different spatial limiters need estimating, and the situation is worse as $N$ is increased. This is another motivation for the simpler closure in Eq.~(\ref{closure3}).}

To minimize the residual, the parameter $\alpha$\ would depend on the unknown angular flux distribution. For simplicity, we fix the $\alpha$~in Eq.~\eqref{closure3}~to a constant. 
Though the central theme is similar to the MLD model in that one adjusts the diffusivity nonlinearly based on the solution, we have not been able to prove that the closure limits the magnitude of $\phi_{N+1}$ to be less than the scalar flux. 

The test results in the following sections demonstrate this modification improves the accuracy in the transients that arise when a majority of the particles in the system have not had a collision. We, therefore, name the model the transient \pn\ closure (\tp{N}).

{
\subsubsection{Closure effects on residual functional derivative}
By introducing the ML\dn\ or T\pn\ closures, the functional derivative in Eq.\ \eqref{func4}\ can be written as:
\begin{subequations}\label{func_tpn}
\begin{align}
\frac{\partial J}{\partial\phi_N}=\frac{(N+1)(N+2)}{2N+3}
\frac{\partial}{\partial\psii{N}}\left[\frac{\partial}{\partial x}\psii{N}\right]
\left(-\frac{1}{v}\partial_t\left(\frac{1}{\st+\nu}\ppz\psii{N}\right)+\frac{\nu}{\st+\nu}\ppz\psii{N}+\partial_x\phi_{N+2}\right)
\end{align}
\begin{equation}
\mathrm{MLD}_N:\quad\nu=\left|\frac{\partial_t\psii{0}}{v\psii{0}}\right|+\left|\frac{\alpha\partial_x\psii{N}}{\psii{0}}\right|
\end{equation}
\begin{equation}
\mathrm{TP}_N:\quad\nu=\left|\frac{\partial_t\psii{0}}{v\psii{0}}\right|+\left|\frac{\alpha\partial_x\psii{0}}{\psii{0}}\right|.
\end{equation}
\end{subequations}
Adding flux or moment limiters does not necessarily minimize the residual functional. In fact, Eq.\ \eqref{func_tpn}\ automatically adjusts the functional derivative based upon the solution. In occasions where the spatial derivative of the solution tends to be large, $1/(\st+\nu)$\ goes to be zero while $\nu/(\st+\nu)$\ limits to  one. Eq.\ \eqref{func_tpn}\ has the limit of \pn's functional derivative. When the solution is smooth and slowly varying in time, $\nu$\ tends to be small and Eq.\ \eqref{func_tpn}\ limits to \dn.\ 
In effect, the closures improve the \dn\, method during transients and preserve the beneficial properties of that closure in the steady limit.
}

\subsubsection{Generalization of T\pn\ models}
{
The form of the T\pn~closure is similar to the Larsen-type flux limited correction to radiation diffusion\cite{Morel:2000vh}\  with an additional time derivative term. The form of a Larsen flux limiter allows the impact of the limiter to be adjusted by making the terms in the closure weighted by a power, rather than using a linear sum.  We can perform the same adjustment to our model by writing
\begin{subequations}
	\begin{equation}
	\psii{N+1}=-\frac{1}{\tilde{\sigma}}\frac{N+1}{2N+3}\ppz\psii{N}
	\end{equation}
	\begin{equation}\label{eq:fluxLim}
	\tilde{\sigma}=\left(\sigma_\mathrm{t}^n+\displaystyle\left|\frac{1}{v\psii{0}}\partial_t\psii{0}\right|^n+\left|\frac{\alpha\partial_x\psii{0}}{\psii{0}}\right|^n\right)^{\frac{1}{n}}
	\end{equation}
\end{subequations}
Typically, the value of $n$ is set to be one or greater, though recent work has demonstrated that there are problems where $n<1$ can give improved solutions \cite{taylor_ans12}. It is then of interest to test the effects from different powers $n$ on our closure.
	}

\subsection{Multi-D extension of T\pn\ closure}
{
The multi-D transport equation with isotropic scattering in Cartesian geometry can be expressed as:
\begin{equation}\label{e:md_trans}
\frac{1}{v}\frac{\partial\psi(\hat{\Omega})}{\partial t}+\hat{\Omega}\cdot\nabla\psi(\hat{\Omega})+\st\psi(\hat{\Omega})=\frac{\sigmas}{4\pi}\int\limits_{4\pi}d\Omega\ \psi(\hat{\Omega})+\frac{Q}{4\pi}.
\end{equation}
In multi-D, \pn\ method is from expanding the angular flux with spherical harmonics functions $Y_l^m(\Omega)$ in angle truncated at Order $N$:
\begin{subequations}
	\begin{equation}\label{e:md_expansion}
	\psi(\hat{\Omega})=\sum\limits_{l=0}^{N}\sum\limits_{|m|=0}^{l}\phi_l^mY_l^m(\hat{\Omega}),
	\end{equation}
	\begin{equation}
	Y_l^m(\hat{\Omega})=Y_l^m(\mu,\varphi)=\begin{cases}
	\sqrt{2}C_l^m\cos(m\varphi)P_l^m(\mu),& m\geq 0\\
	\sqrt{2}C_l^{|m|}\sin(|m|\varphi)P_l^{|m|}(\mu),& m>0\\
	\end{cases},
	\end{equation}
\end{subequations}
where \[C_l^m=\displaystyle\sqrt{\frac{(2l+1)}{4\pi}\frac{(l-m)!}{(l+m)!}}.\] In a similar Galerkin procedure as in Section\ \ref{s:derive}\, one uses the expansion in Eq.\ \eqref{e:md_expansion}\ in Eq.\ \eqref{e:md_trans}\ and operate on the transport equation with $\int\limits_{4\pi}d\Omega\ {Y}_l^m(\cdot)$\ for all the angular basis functions. The result is
\begin{subequations}
	\begin{equation}
	\frac{1}{v}\frac{\partial\vec{\phi}}{\partial t}+\sum\limits_{\zeta=x,y,z}\partial_\zeta\mathbf{A}_\zeta\vec{\phi}+\mathbf{\Sigma}_\mathrm{r}\vec{\phi}=\vec{Q}
	\end{equation}
	\begin{equation}
	\mathbf{A}_{\zeta,l,m}^{l',m'}=\int\limits_{4\pi}d\Omega\ {Y}_{l}^{m}(\hat{\Omega})\Omega_\zeta Y_{l'}^{m'}(\hat{\Omega})
	\end{equation}
	\begin{equation}
	\vec{\phi}=\left(\phi_0^0,\phi_1^{-1},\cdots,\phi_1^1,\cdots,\phi_N^{-N},\cdots,\phi_N^N\right)^\top,\quad\vec{Q}=(Q,0,\cdots,0)^\top
	\end{equation}
	\begin{equation}
	\mathbf{\Sigma}_\mathrm{r}=\mathrm{diag}(\st-\sigmas,\st,\cdots,\st)
	\end{equation}
\end{subequations}
Due to symmetry, there are $(N+1)(N+2)$\ moments in 3D and $(N+1)(N+2)/2$\ relevant moments in 2D.
	}

{
\subsubsection{\dn\ equations}
The \dn\ model is identical to \pn\ up to the $(N-1)^\mathrm{th}$\ moment equations. Dropping off the time derivative terms of the $(N+1)^\mathrm{th}$\ moment equations of P$_{N+1}$\ system, one can easily find:
	\begin{equation}\label{dn_2d}
	\phi_{N+1}^m=-\frac{1}{\st}\sum\limits_{\zeta=x,y,z}\sum\limits_{l',m'}\mathbf{A}_{\zeta,N+1,m}^{l',m'}\partial_\zeta\phi_{l'}^{m'},\quad |m|\leq N+1
	\end{equation} 
Plugging Eq.\ \eqref{dn_2d}\ into the relevant moment equations up to Order N will then lead to the \dn\ system. Note that Eq.\ \eqref{dn_2d}\ illustrates that \dn\ is equivalent to adding a diffusive correction to the P$_{N-1}$\ system\cite{levermoredn}.
\subsubsection{T\pn\ closure}
It is straightforward to extend modify the \dn\ model to be T\pn\ by adding a correction term $\nu$\ to the denominator of Eq.\ \eqref{dn_2d}:
	\begin{subequations}
		\begin{equation}\label{tpn_2d}
		\phi_{N+1}^m=-\frac{1}{\st+\nu}\sum\limits_{\zeta=x,y,z}\sum\limits_{l',m'}\mathbf{A}_{\zeta,N+1,m}^{l',m'}\partial_\zeta\phi_{l'}^{m'},\quad |m|\leq N+1
		\end{equation}
		\begin{equation}\label{viscosity}
		\nu\equiv\left(\left|\frac{\partial_t\phi_0^0}{v\phi_0^0}\right|+\alpha\frac{\|\nabla\phi_0^0\|}{|\phi_0^0|}\right).
		\end{equation}
	\end{subequations}
	}

\section{Numerical Details} 
{
The 1D ML\dn\ and T\pn\ closures are implemented with the diamond difference for spatial discretization and a semi-implicit scheme as detailed below. At present only T\pn\ closure is extended to multi-D applications with discontinuous Galerkin (DG) finite element method in space and semi-implicit scheme in time.
	}
\subsection{1D implementation}
For our closures the highest order moment we keep in our system is $N$ with $N$ even.  With $N$ even, there are $N$\ first-order PDEs and one second-order PDE.  This requires $N+2$ total boundary conditions or $N/2 + 1$ conditions on each boundary in 1-D. We can use the standard Marshak conditions in this case where on the left boundary we satisfy 
\begin{equation}\label{e:bdy}
\int\limits_{0}^{1}\!d\mu\,\psi_{\mathrm{inc}}^\mathrm{L}(\mu)P_l(\mu)=\sum\limits_{i=0}^{N}c_i\psii{i}-\frac{c_{N+1}}{\sigma_\mathrm{t}+\displaystyle\left|\frac{\partial_t\psii{0}}{v\psii{0}}\right|+\alpha\left|\frac{\partial_x\psii{k}}{\psii{0}}\right|}\frac{N+1}{2N+3}\ppz\psii{N}, \qquad l=1,3,..,{N+1},,
\end{equation}
where $k$ is equal to $0$ or $N$, and $\displaystyle c_i=\int_{0}^1d\mu~P_l(\mu)P_i(\mu)$. The conditions at the right boundary are the same except the integral is over $\mu\in[-1,0]$. 

For a spatial discretization we use the diamond difference method with unknowns that live at cell edges. For a uniform mesh with cell width $h$, the semi-discrete equations become 
\begin{subequations}\label{eqs:DD}
	\begin{multline}
	\frac{h}{v}\partial_t\psii{l,i}+\frac{l}{2l+1}\left(\psii{l-1,i+1/2}-\psii{l-1,i-1/2}\right)+\frac{l+1}{2l+1}\left(\psii{l+1,i+1/2}-\psii{l+1,i-1/2}\right)\\
	+h\left(\sigma_{\mathrm{t},i}-\sigma_\mathrm{s,i}\delta_{0,l}\right)\psii{l,i}=Q_{l,i}\delta_{0,l}h,\quad l=0,\cdots,N-1
	\end{multline}
	\begin{multline}
	\frac{h}{v}\partial_t\psii{N,i}+\frac{N}{2N+1}\left(\psii{N-1,i+1/2}-\psii{N-1,i-1/2}\right)+h\sigma_{\mathrm{t},i}\psii{N,i}\\
	-\frac{N(N+1)}{(2N+1)(2N+3)}\left(\frac{\partial_x\psii{N,i+1/2}}{\tilde{\sigma}_{i+1/2}}-\frac{\partial_x\psii{N,i-1/2}}{\tilde{\sigma}_{i-1/2}}\right)=0
	\end{multline}
	\begin{equation}
	\tilde{\sigma}_{i}=\sigma_{\mathrm{t,}i}+\left|\frac{\partial_t\psii{0,i}}{v\psii{0,i}}\right|+\alpha\left|\frac{\psii{k,i+1/2}-\psii{k,i-1/2}}{h\psii{0,i}}\right|,
	\end{equation}
\end{subequations}
where $k$ is either $0$ or $N$, and
\begin{subequations}
	\begin{equation}
	\partial_x\psii{l,i+1/2}=\frac{\psii{l,i+1}-\psii{l,i}}{h},
	\end{equation}
	\begin{equation}
	\psii{l,i}=\frac{1}{2}\left(\psii{l,i+1/2}+\psii{l,i-1/2}\right),\quad l=0,\cdots,N,
	\end{equation}
	\begin{equation}
	\tilde{\sigma}_{i+1/2}=\frac{1}{2}\left(\tilde{\sigma}_{i+1}+\tilde{\sigma}_{i}\right).
	\end{equation}
\end{subequations}

The time discretization we use is semi-implicit that we evaluate all terms in  Eq.~(\ref{eqs:DD})\ at time level $n+1$ (i.e., backward Euler) except $\tilde{\sigma}$, which is evaluated explicitly at level $n$. This makes each time step a linear solve.  If we implicitly update $\tilde{\sigma}$, each step would require a nonlinear solve. 
\subsection{2D T\pn\ implementation}
{
Previously, \dn\ has been discretized in space by the streamline diffusion continuous finite element method\cite{levermoredn}\ and finite volume method\cite{p3qs,cory_hauck_closures}\ in space.\ We choose a variant of the discontinuous Galerkin (DG) finite element method  in this work mainly for its preservation of the asymptotic diffusion limit. In particular, we apply the local DG (LDG)\ method, which was developed for time dependent convection-diffusion equation\cite{cockburn_ldg}:\
\begin{equation}\label{conv_dif}
\frac{\partial u}{\partial t}+\nabla\cdot\mathrm{F}(u)+\nabla a(x,y,z)\nabla u=0.
\end{equation}
In the LDG method, one introduces an auxiliary variable $\vec{q}$,\ such that Eq.\ \eqref{conv_dif}\ can be rewritten as:
\begin{subequations}\label{ldg}
\begin{equation}\label{ldg1}
\frac{\partial u}{\partial t}+\nabla\cdot\mathrm{F}(u)+\nabla\cdot\vec{q}=0,
\end{equation}
\begin{equation}
\vec{q}=a(x,y,z)\nabla u.
\end{equation}
\end{subequations}
	}

{
	}
\section{Numerical results}
The 2D T\pn\ closure is implemented with the C\texttt{++} open source finite element library deal.II\cite{dealii82}.\ The results for the plane source problem, two-beam problem and the Reed's problem will be presented for 1D closures and 2D T\pn\ test results will be presented with line source problem.

\subsection{Plane source test problem}
\begin{figure}[ht!]
	\begin{subfigure}{.5\textwidth}
		\centering
		\hspace*{-1cm}\includegraphics[width=1.\linewidth]{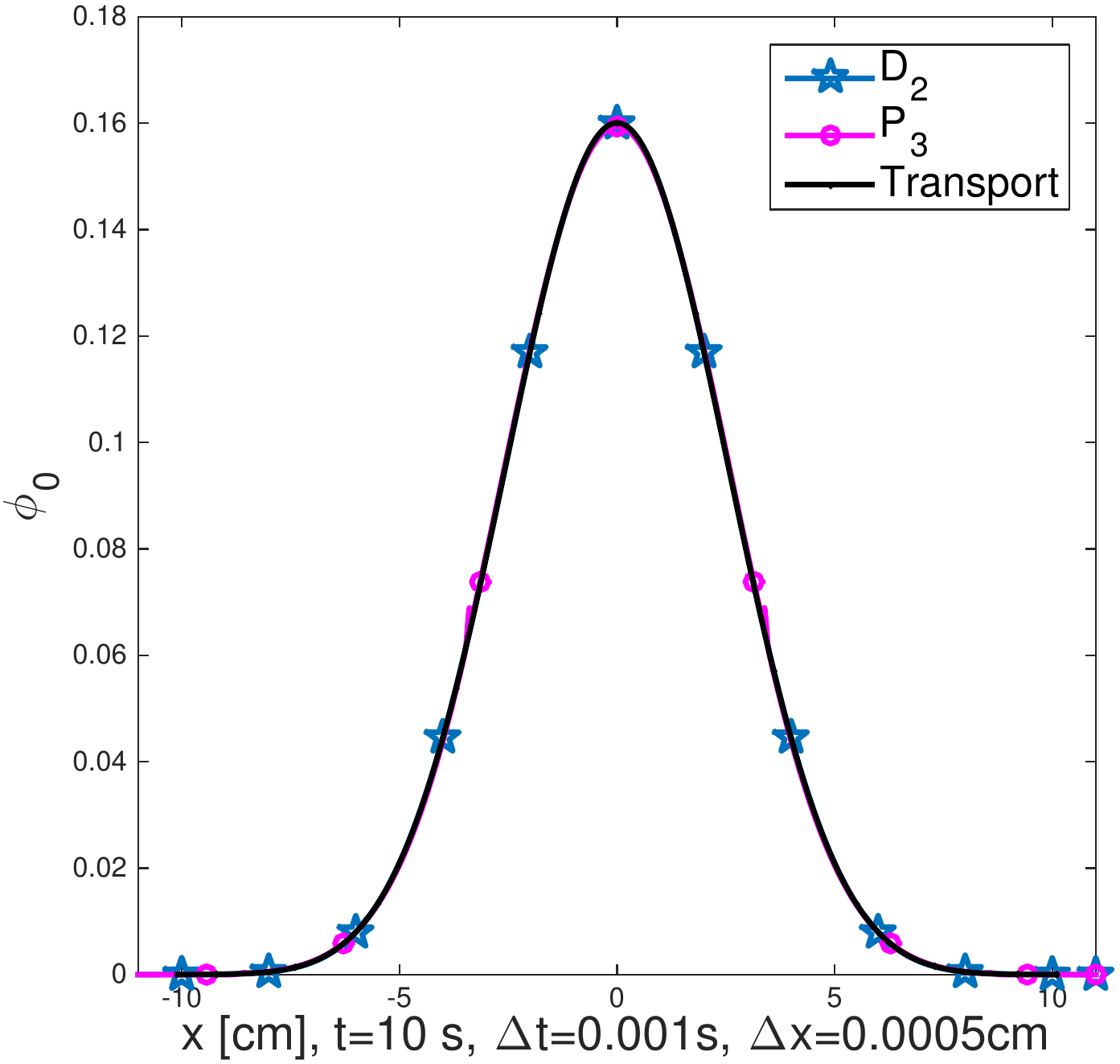}
		\caption{Late time solutions ($t=10$)}
		\label{f:gd0}
	\end{subfigure}
	~
	\begin{subfigure}{.5\textwidth}
		\centering
		\includegraphics[width=1.\linewidth]{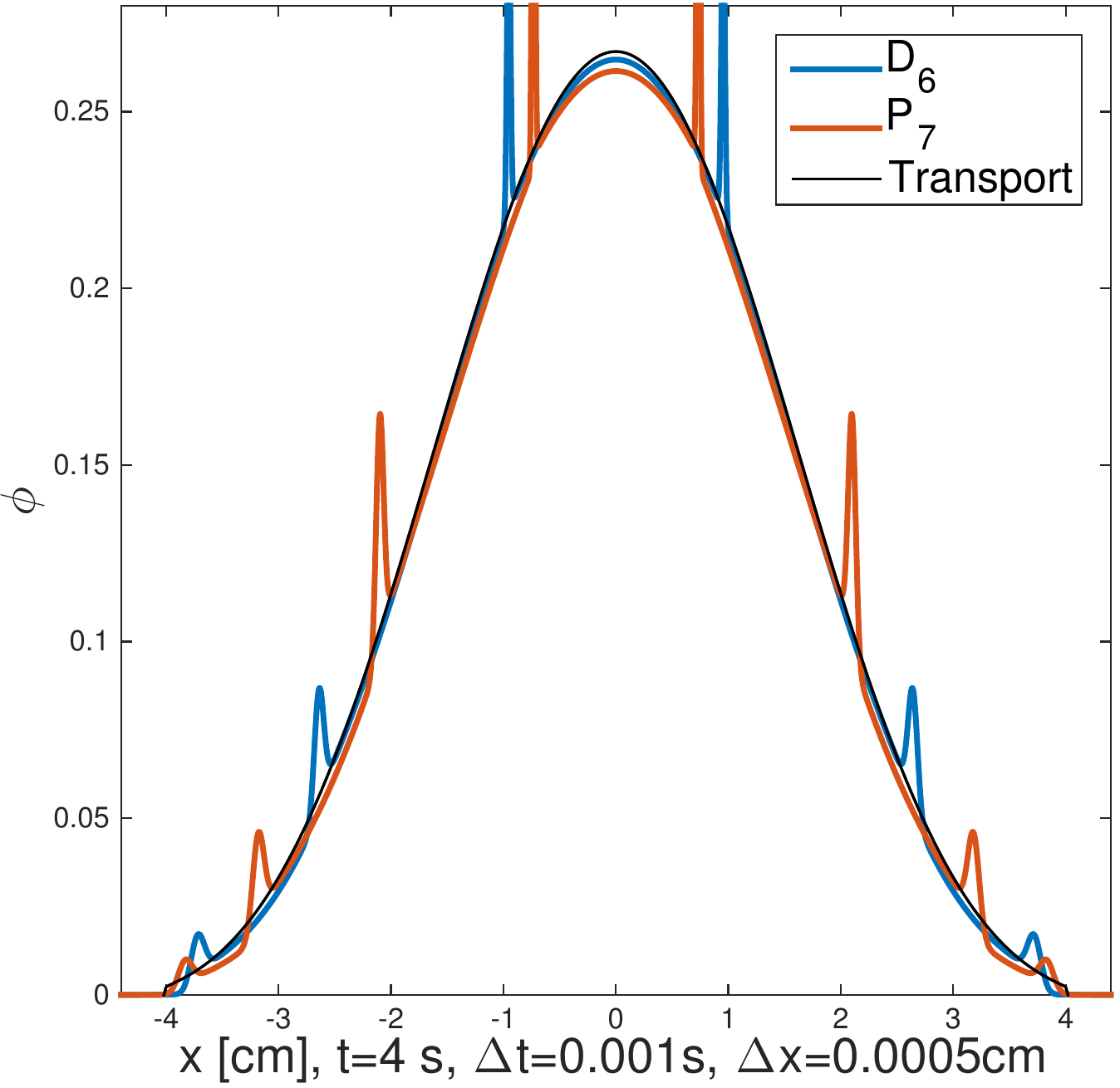}
		\caption{Solutions at an earlier time ($t=4$)}
		\label{f:bd0}
	\end{subfigure}
	\caption{Examples of \pn\ and \dn\ in plane source problem. Notice that early in time the discrete wave speeds in the \pn and \dn~solutions.}
	
\end{figure}

The medium in the plane source problem is a pure scatterer ($\st =\sigma_\mathrm{s}=1$).~At time $t=0$,~there is a pulsed source in the middle of an infinite slab. The initial condition is
\begin{equation}
\psi(z,\mu,0)=\frac{\delta(z)}{2}.
\end{equation}
{
An analytic solution to the transport equation for this problem is available in the benchmark suite AZURV1\cite{ganapol}.
The solution has a wavefront at $z=\pm vt$. The number of particles in the wavefront decays over time so that after enough time the wavefront has a negligible magnitude. Therefore, late in time the solution is a smooth due to the scattering of particles from the initial pulse. Also, both the \dn\: and \pn\: methods approximate the transport solution well at late times (e.g., $t=10$ in Figure \ref{f:gd0}), whereas early on in the transient neither can capture the analytic solution. This is predicted by the analysis in Section\ \ref{s:above}\ because early in time the spatial and time derivatives of the solution are not small. At $x=0$ the D$_6$ solution is closer to the analytic solution than the P$_7$. 
	}

{
In the solution at earlier times (see Fig.~\ref{f:bd0}) there are spikes that are the numerical representation of waves of uncollided particles. Since the time dependent \pn\ system is a hyperbolic wave equation system, particles moves in several discrete wave speeds. The consequence is that the solutions will have $N+1$~spikes,~analytically represented by a Dirac delta function. These artifacts from the \pn\ (and \dn) discretization are known as wave effects\cite{brunner_app_rad_trans}.
	}

\subsubsection{Comparison of ML\dn\ and linear closures}
In the results below, unless otherwise noted, we use a value of $\alpha = 2/3$. Later, we discuss this choice.

\begin{figure}[ht!]
	\begin{subfigure}{.5\textwidth}
		\centering
		\hspace*{-1cm}\includegraphics[width=1.\linewidth]{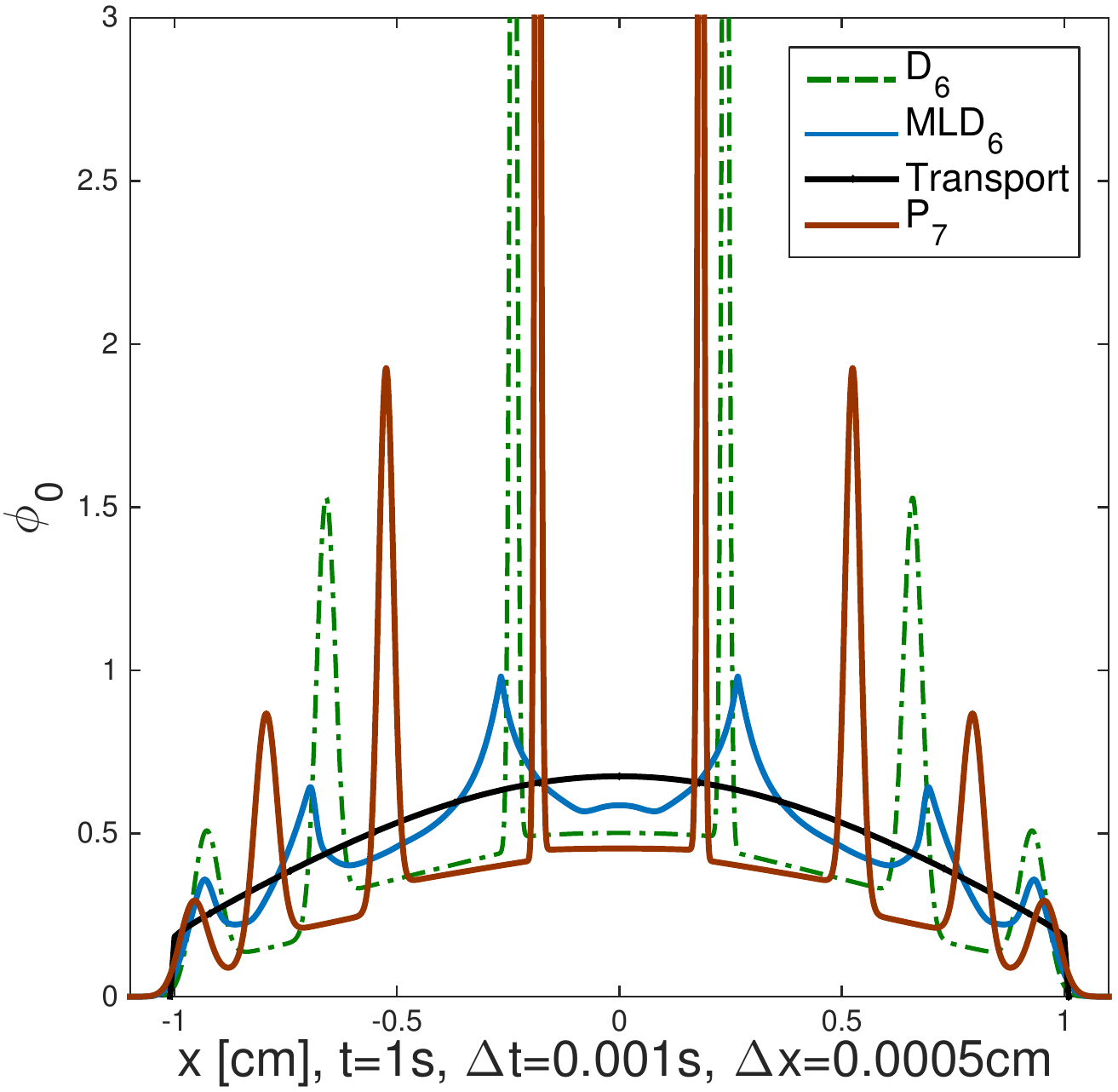}
		\caption{MLD$_6$\ results at 1s in plane source problem.}
		\label{f:mld4}
	\end{subfigure}
	~
	\begin{subfigure}{.5\textwidth}
		\centering
		\includegraphics[width=1.\linewidth]{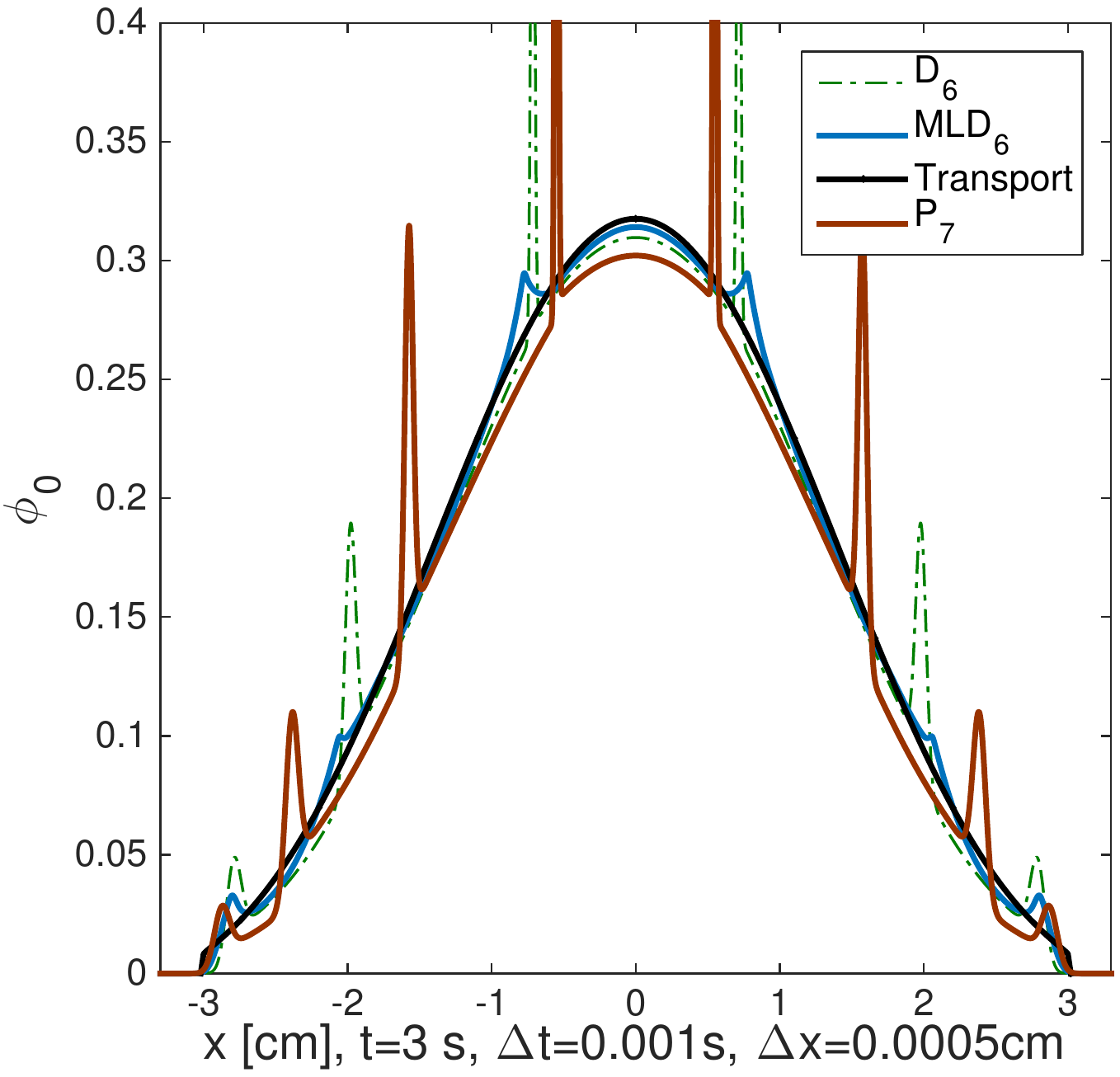}
		\caption{MLD$_6$\ results at 3s in plane source problem.}
		\label{f:mld6}
	\end{subfigure}
	\caption{ML\dn~and linear closure~solutions to the plane source problem at different times.}
	\label{mls}
\end{figure}

In Figure  \ref{mls} we compare MLD and the diffusive closure on the plane source problem. At an early time, Figure \ref{f:mld4}, the wave effects are greatly reduced in the MLD$_6$ model relative to D$_6$\ and P$_7$.\ Furthermore, the solution away from the waves is much closer to the transport solution. At later time, Figure \ref{f:mld6}, the wave effects in P$_7$ and D$_6$ are still present whereas the MLD$_6$ solution has the overall shape of the transport solution with small oscillations near the D$_6$ waves. 

\subsubsection{Comparison of ML\dn\ and \tp{N}}
\begin{figure}[ht!]
	\begin{subfigure}{.5\textwidth}
		\centering
		\hspace*{-1cm}\includegraphics[width=1.\linewidth]{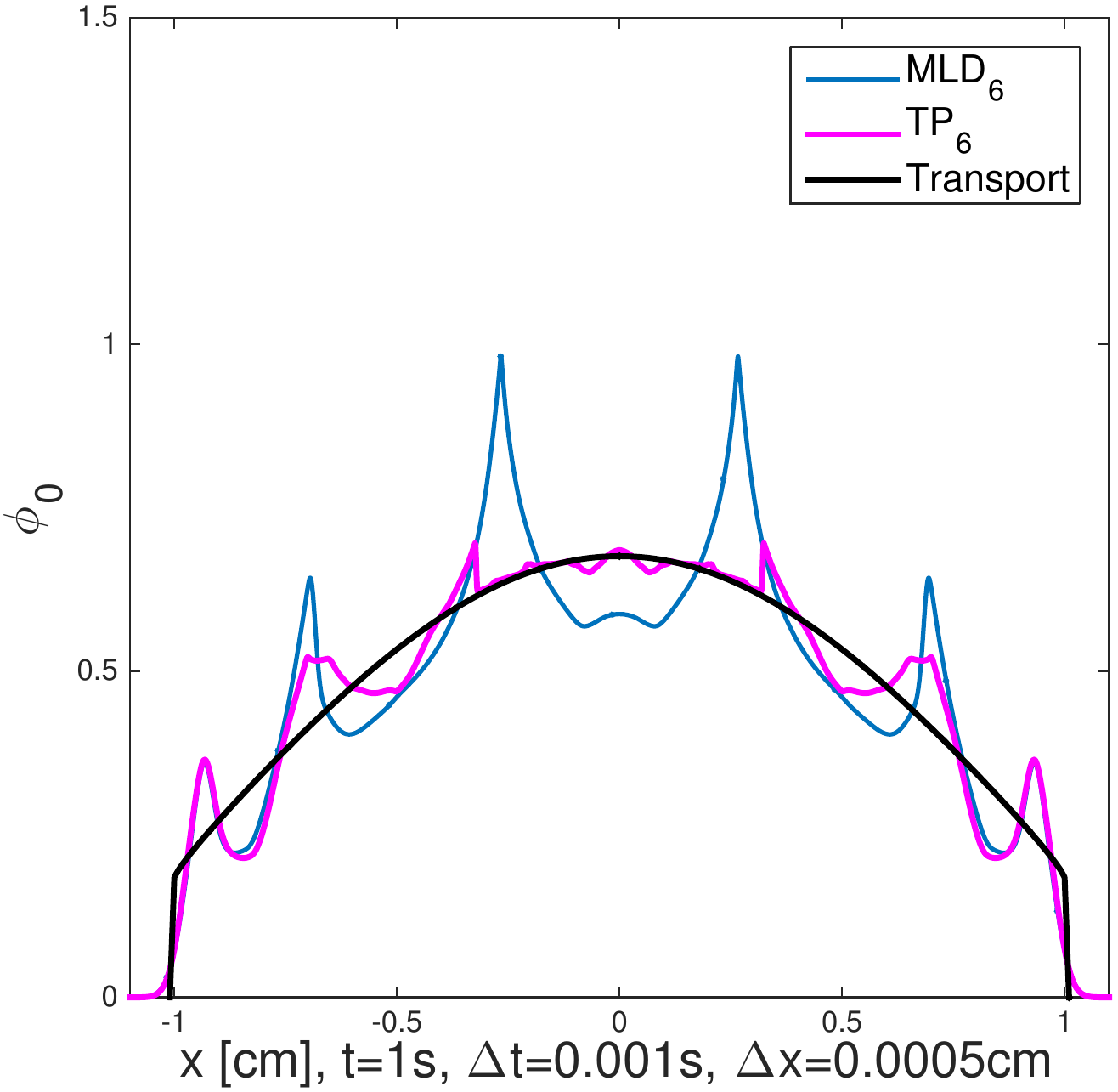}
		\caption{MLD$_6$\ and TP$_6$\ results.}
		\label{f:mlfl7}
	\end{subfigure}
	~
	\begin{subfigure}{.5\textwidth}
		\centering
		\includegraphics[width=1.\linewidth]{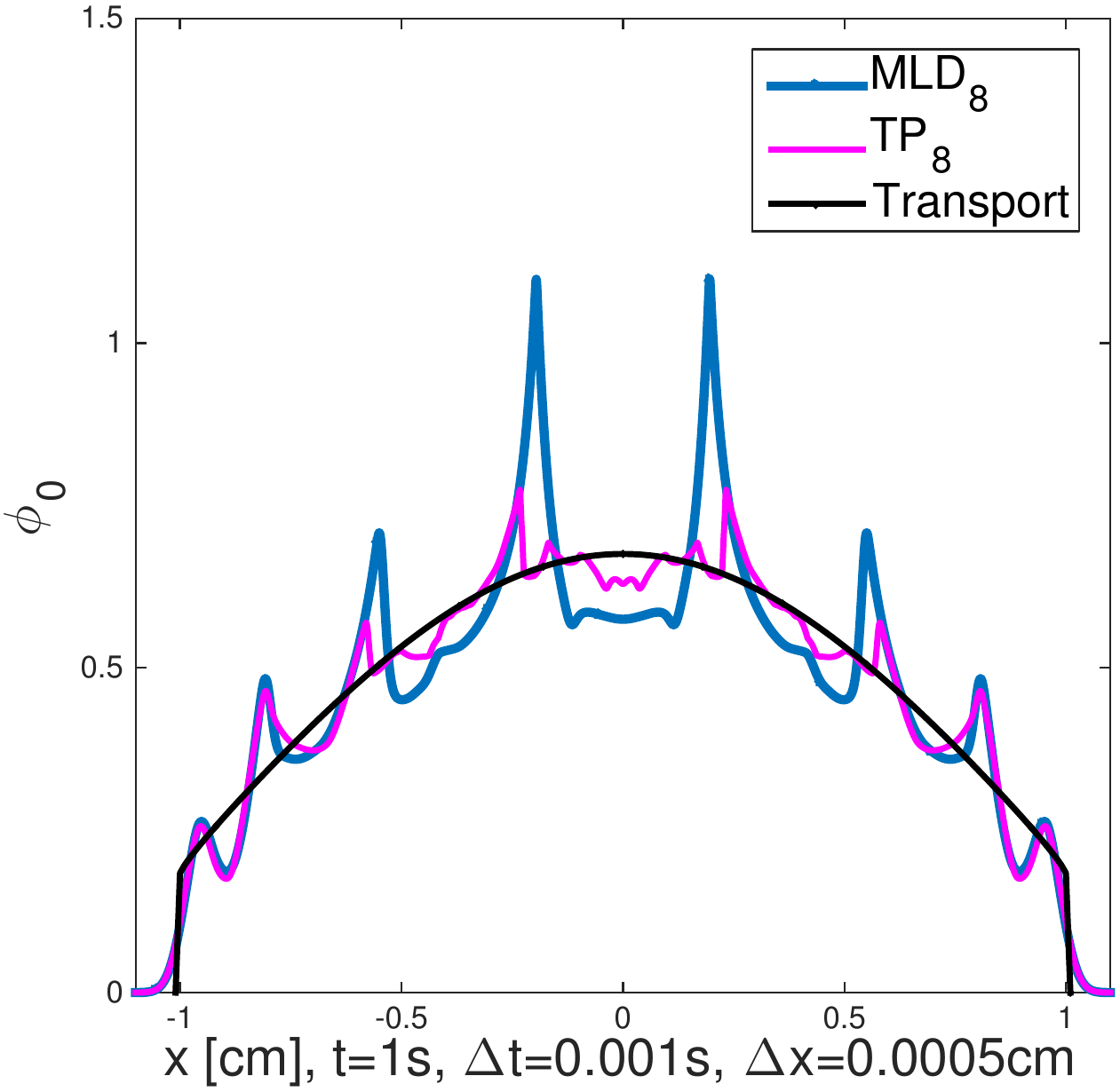}
		\caption{MLD$_8$\ and TP$_8$\ results.}
		\label{f:mlfl9}
	\end{subfigure}
	\caption{ML\dn\ and T\pn\ comparison at 1 s.}
	\label{mlfls}
\end{figure}

We next compare the two models developed in this paper.  At $t = 1$s\ in the plane source problem, as shown in Figure \ref{mlfls}, with both $N=6$ and $8$, the T\pn~model gives results closer to the transport solution than the ML\dn~model. 

\begin{figure}[ht!]
	\begin{subfigure}{.5\textwidth}
		\centering
		\hspace*{-1cm}\includegraphics[width=1.\linewidth]{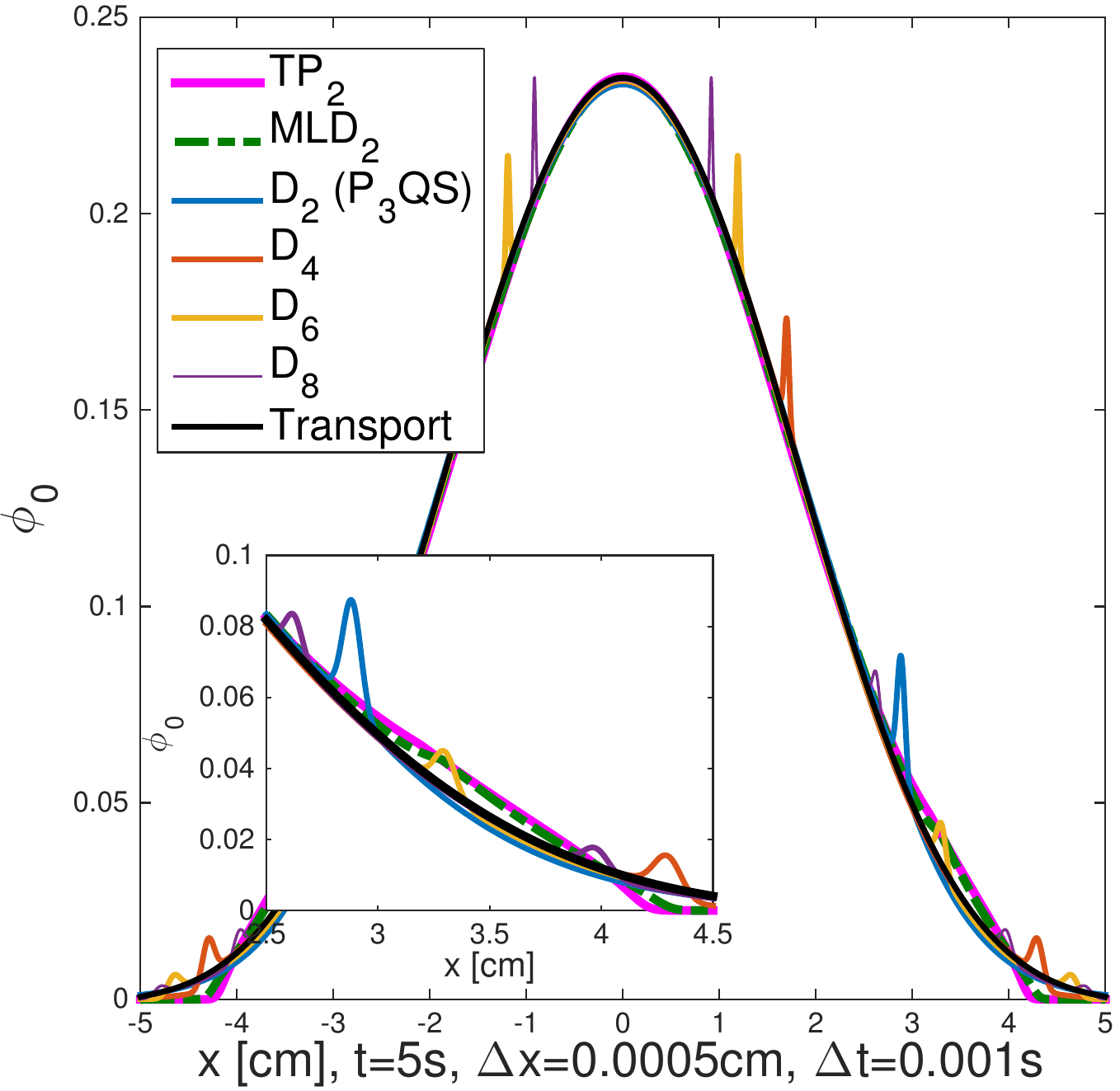}
		\caption{MLD$_2$\ and TP$_2$\ results.}
		\label{f:flml3}
	\end{subfigure}
	~
	\begin{subfigure}{.5\textwidth}
		\centering
		\includegraphics[width=1.\linewidth]{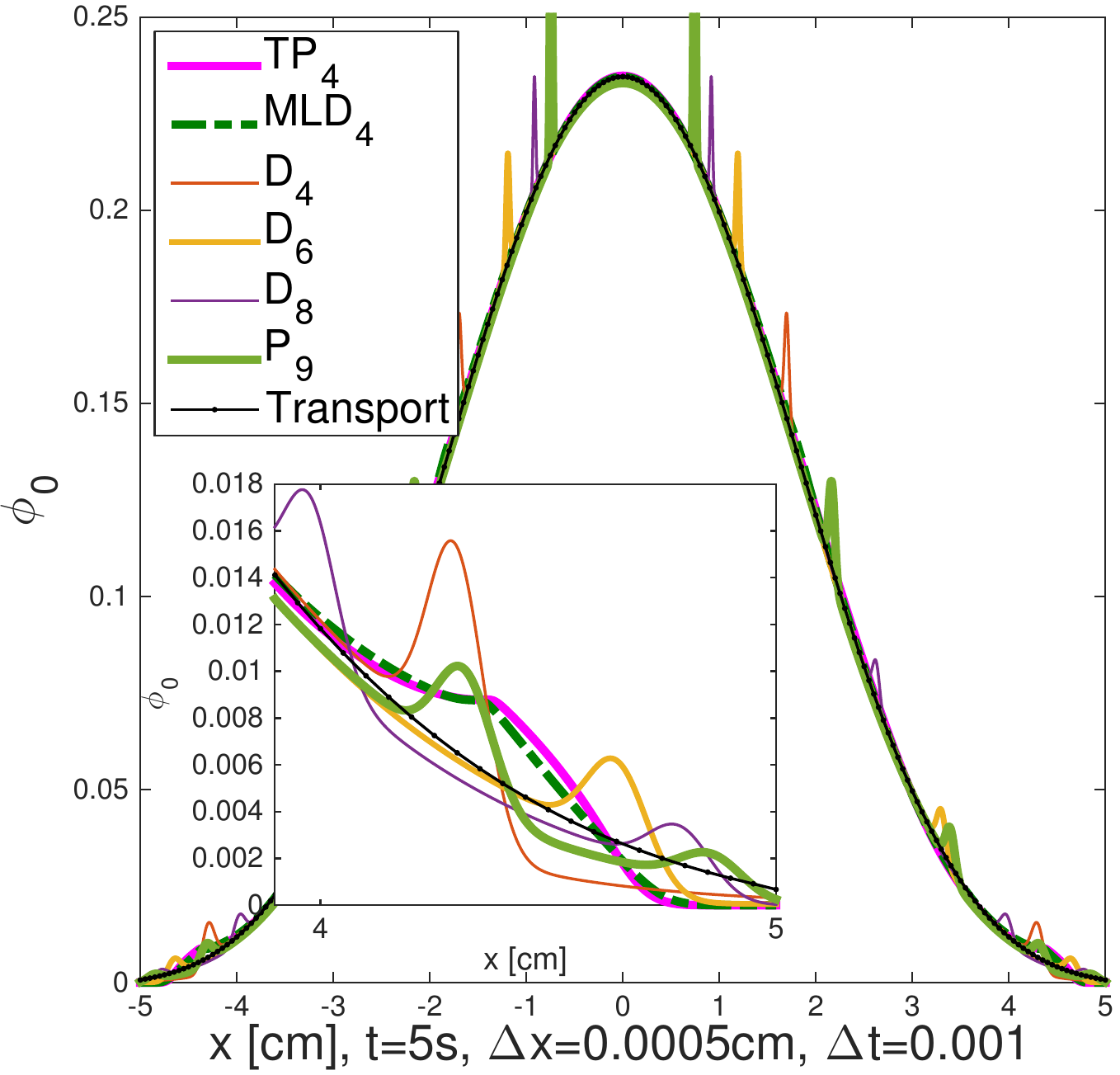}
		\caption{MLD$_4$\ and TP$_4$\ results.}
		\label{f:flml5}
	\end{subfigure}
	\caption{ML\dn\ and T\pn\ comparison at 5s.}
	\label{mlfls2}
\end{figure}

At 5s after the pulse, it is seen that in Figure\ \ref{mlfls2},\ both closures do not produce artificial waves in the solution to the degree that \dn~or \pn~solutions do. The  MLD$_N$\ and TP$_N$\ results basically agree to the transport solution in the middle except the solution near the wavefronts in the $\pm 5$\ cm. At the wavefront none of these methods captures the solution correctly.
In summary, both the ML\dn~and T\pn\ closures effectively damp the unphysical modes (large spikes), which leads to relatively accurate solutions for the transport problem during short-time transients. Moreover, on every problem we have tested, the T\pn~ method was superior to the ML\dn~method.  Henceforth, we will focus on this method.

\subsubsection{The impact of spatial and temporal terms in the model}
\begin{figure}[ht!]
	\begin{subfigure}{.5\textwidth}
		\centering
		\hspace*{-1cm}\includegraphics[width=1.\linewidth]{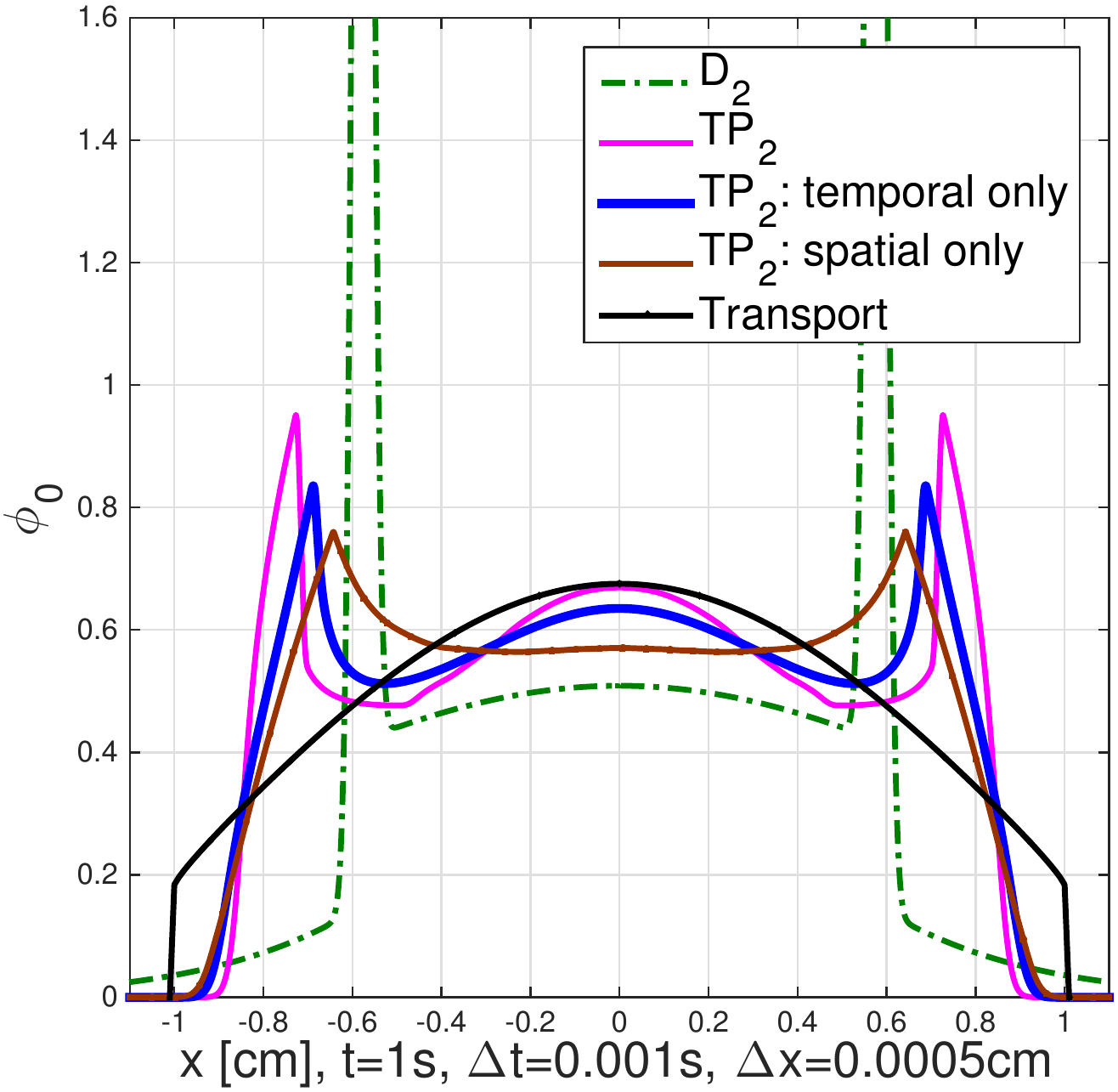}
		\caption{TP$_2$ results at $t = 1$ s}
		\label{f:fl3limiter}
	\end{subfigure}
	~
	\begin{subfigure}{.5\textwidth}
		\centering
		\includegraphics[width=1.\linewidth]{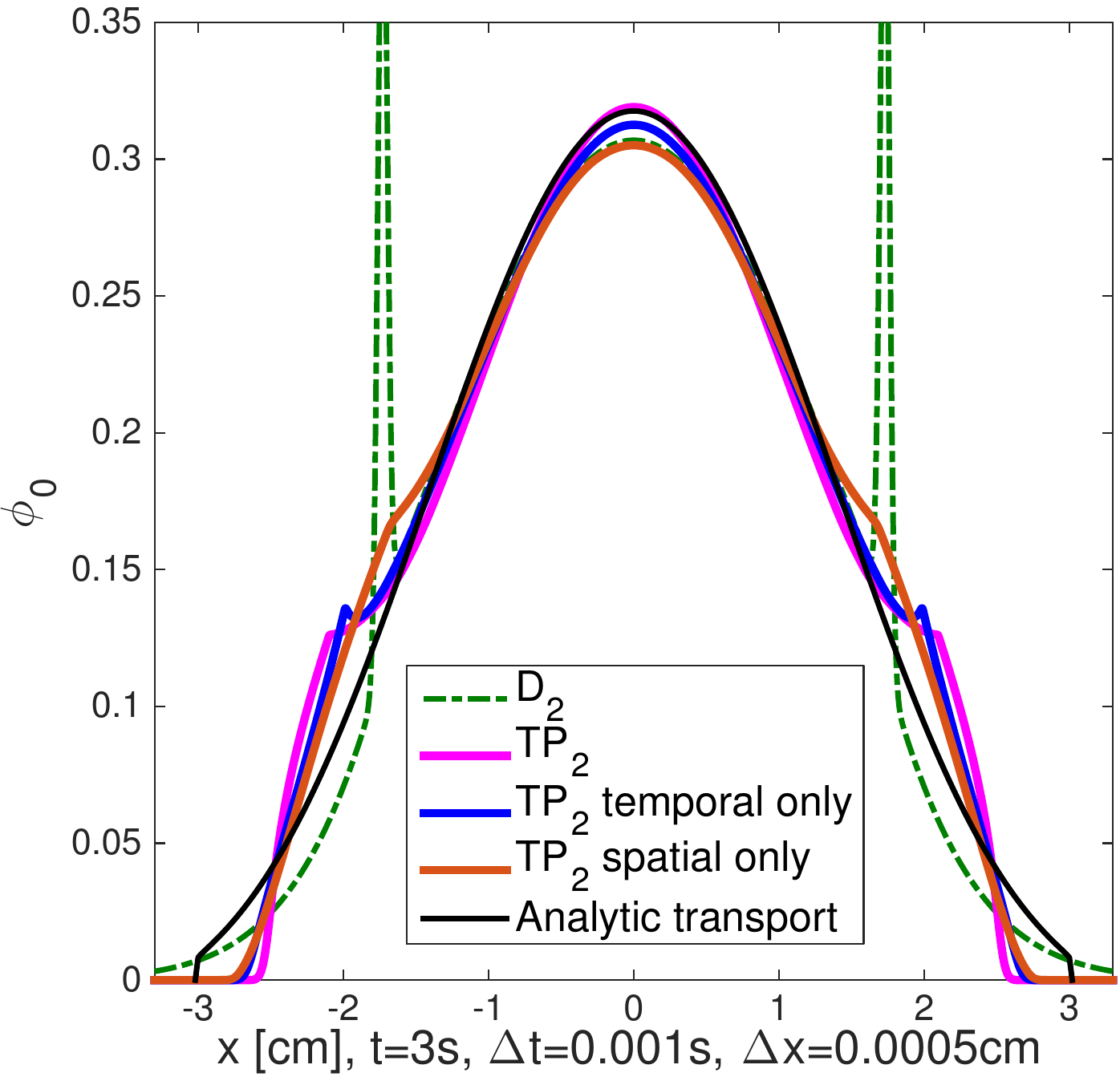}
		\caption{TP$_2$ results at $t = 3$ s}
		\label{f:fl5limiter}
	\end{subfigure}
	\caption{Illustration of the impact of the spatial and temporal derivative terms in the T\pn~closure.}
	\label{f:limiters}
\end{figure}

To further investigate the importance of different terms in the closure, we individually turn on/off different derivatives in the closure. It is observed that, at 1\ s in the plane source problem in Figure \ref{f:fl3limiter} using only the spatial derivative term in the closure makes the solution flat in the middle and, as a result, too low. On the other hand, merely using the temporal derivative terms retains a better flux profile in the slab center, while the artificial spikes are not yet dampened effectively as with spatial limiter at later times in Figure \ref{f:fl5limiter}. Note that the moving modes propagate further than those with the spatial limiter. Therefore, we conclude that both derivative terms in the closure contribute to the accuracy of the model. 


\subsubsection{Impact from Power $n$\ of T\pn\ models}
\begin{figure}[ht!]
\centering
\includegraphics[width=.5\linewidth]{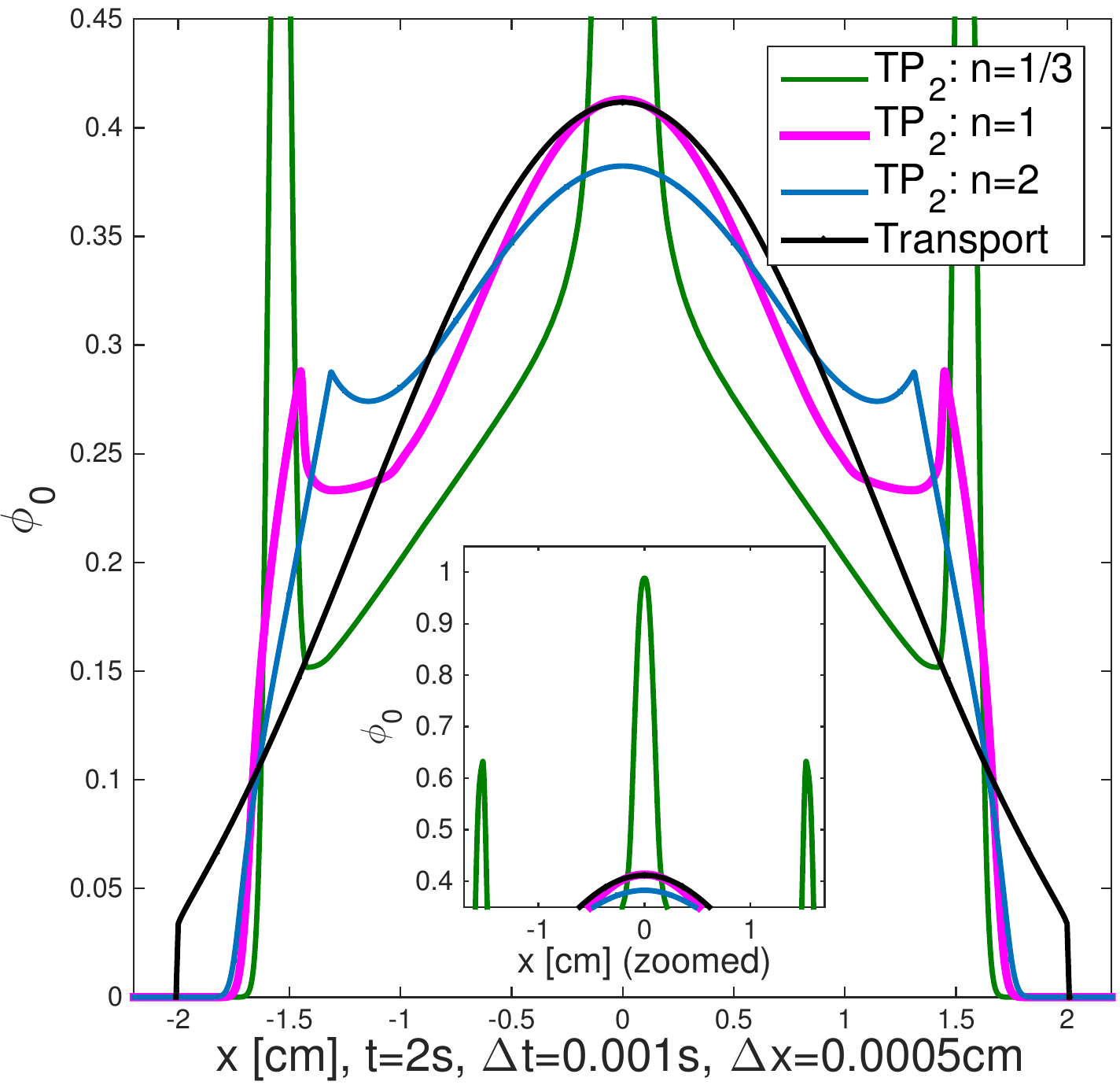}
\caption{Effects from different power $n$ on the in Eq.~(\ref{eq:fluxLim}).}
\label{f:tp3n}
\end{figure}

 It is observed that for low order T\pn\ approximations, varying the power $n$ does adjust the dissipation in the solution. As illustrated in Figure\ \ref{f:tp3n}, the originally proposed value, $n=1$\ retains the correct value near $x=0$. Simultaneously, $n=2$\ makes the solution flatter. On the other hand, reducing $n$\ to $1/3$ amplifies the dampened spikes and makes the solution more similar to the even \pn\ flux profile in that it has a stationary mode at $x=0$. It would suggest small powers should be avoided. Yet, all solutions agree with each other when the transient is passed.
We have also observed that with increasing $N$\ the solution becomes less sensitive to the power $n$ for $n>1$.

\subsubsection{Spatial derivative coefficient $\alpha$}
In our initial derivation of the T\pn~model we surmised that the value of $\alpha$ should be between $1/3$ and 1 because it appears in a similar way to the coefficients of the \pn~Jacobian. Since these coefficients range from $1/3$ through 1, we therefore used the median value $2/3$. 

\begin{figure}[ht!]

	\centering
	\includegraphics[width=.5\linewidth]{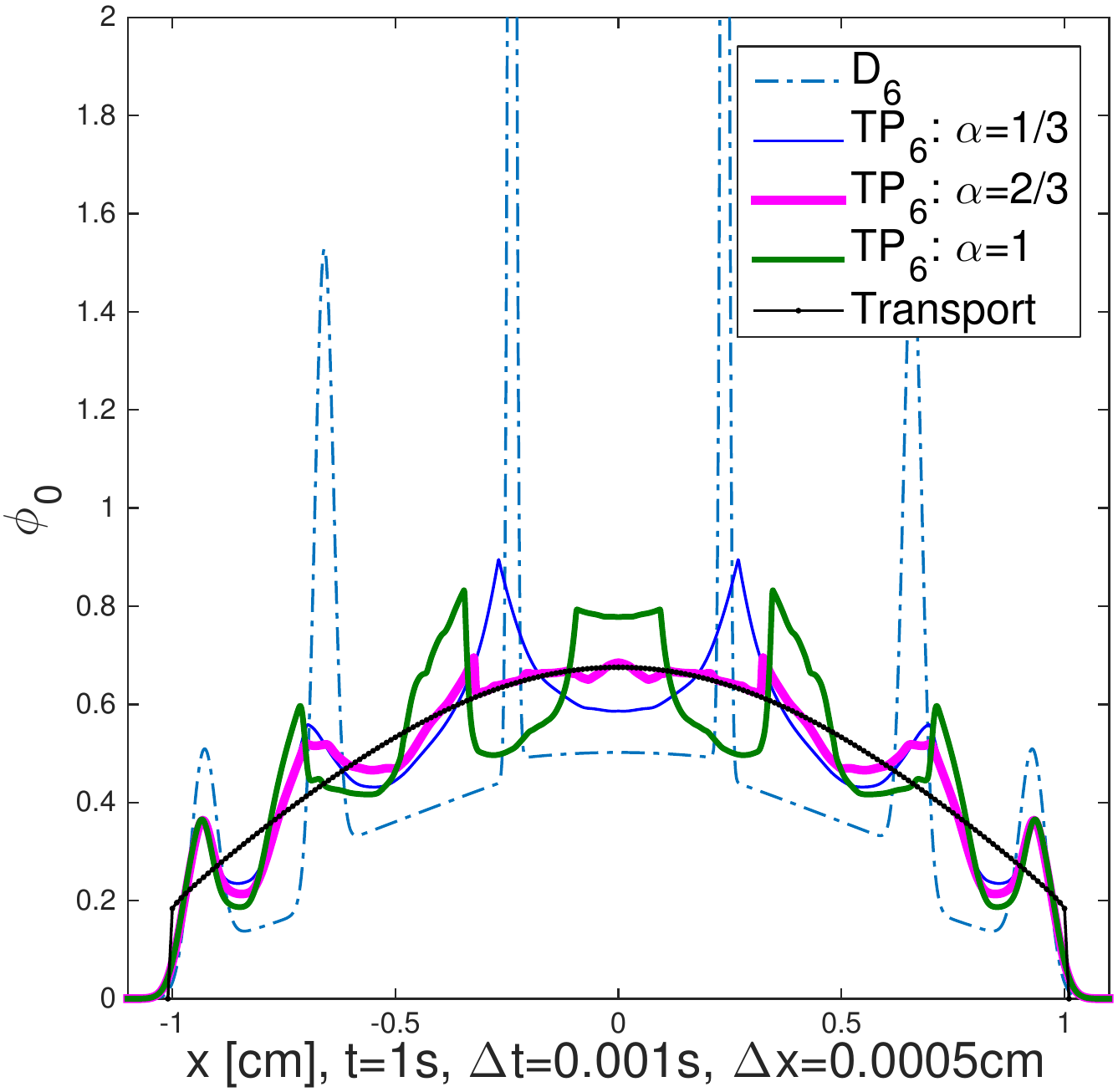}
	\caption{Effects from different coefficients $\alpha$ in the closure. With $\alpha =0 $ the spatial derivative of the scalar flux has no impact on the closure.}
	\label{f:tp6coef}
\end{figure}

We present a limited parameter study for $\alpha$  in Figure\ \ref{f:tp6coef}. Therein, the flux profiles vary in several respects. With the smallest value shown, $\alpha=1/3$, the solution is much closer to the \dn~solution: the solution is too low in the middle, and the wave effects are amplified. On the other hand, increasing $\alpha$\ to 1 appears to amplify and spread the waves in the solution in addition to increasing the solution near $x=0$ too much.

Compared with 1/3 and 1, $2/3$\ provides the most accurate and least oscillatory result among the three choices. We have performed more  studies using many more values of $\alpha$ and found that  $\alpha$ values of 0.5 through 0.7 are comparable to the 2/3 solution.

\subsection{Two-beam problem}

\begin{figure}[ht!]
	\begin{subfigure}{.5\textwidth}
		\centering
		\hspace*{-1cm}\includegraphics[width=1.\linewidth]{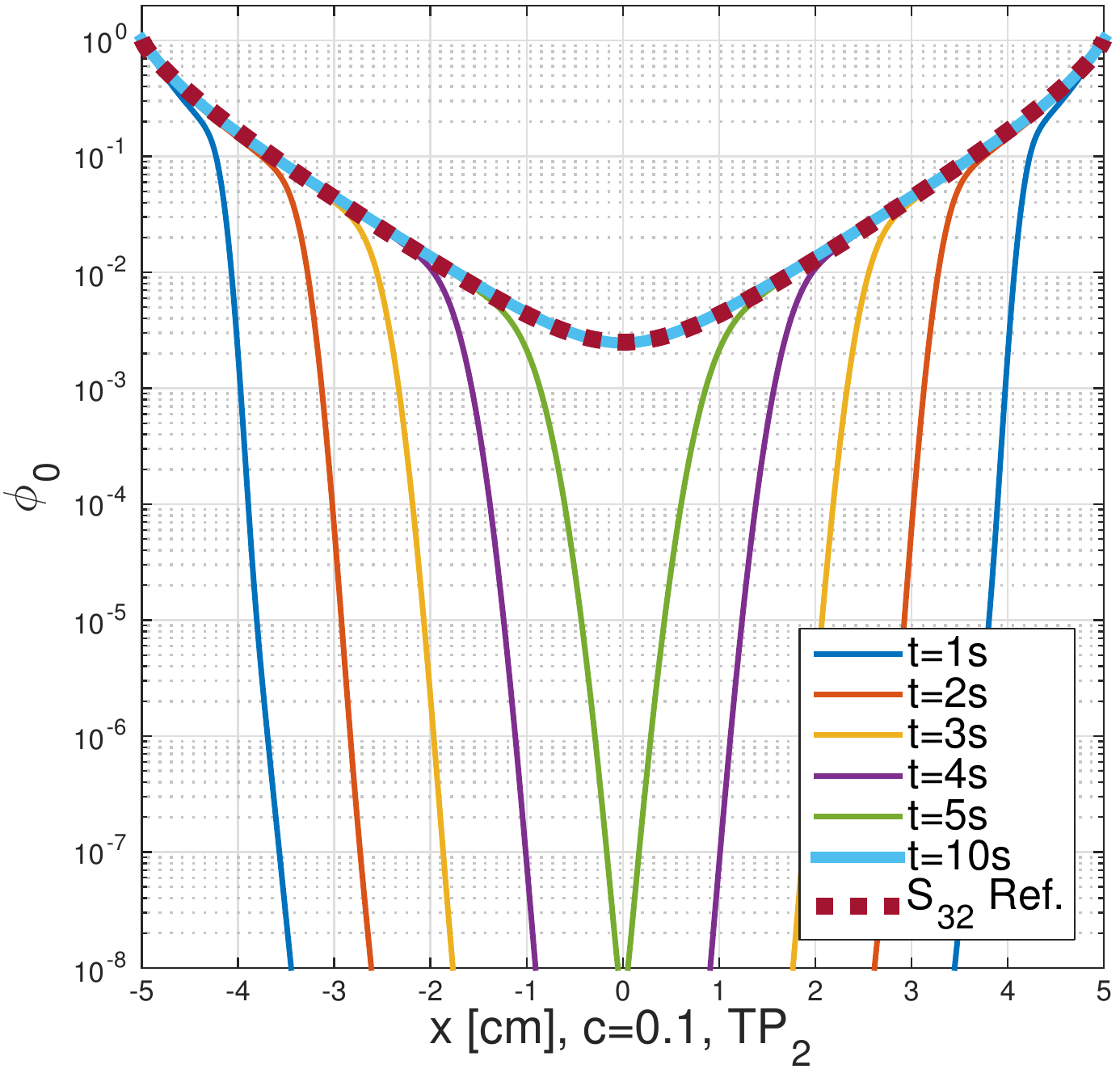}
	\end{subfigure}
	~
	\begin{subfigure}{.5\textwidth}
		\centering
		\includegraphics[width=1.\linewidth]{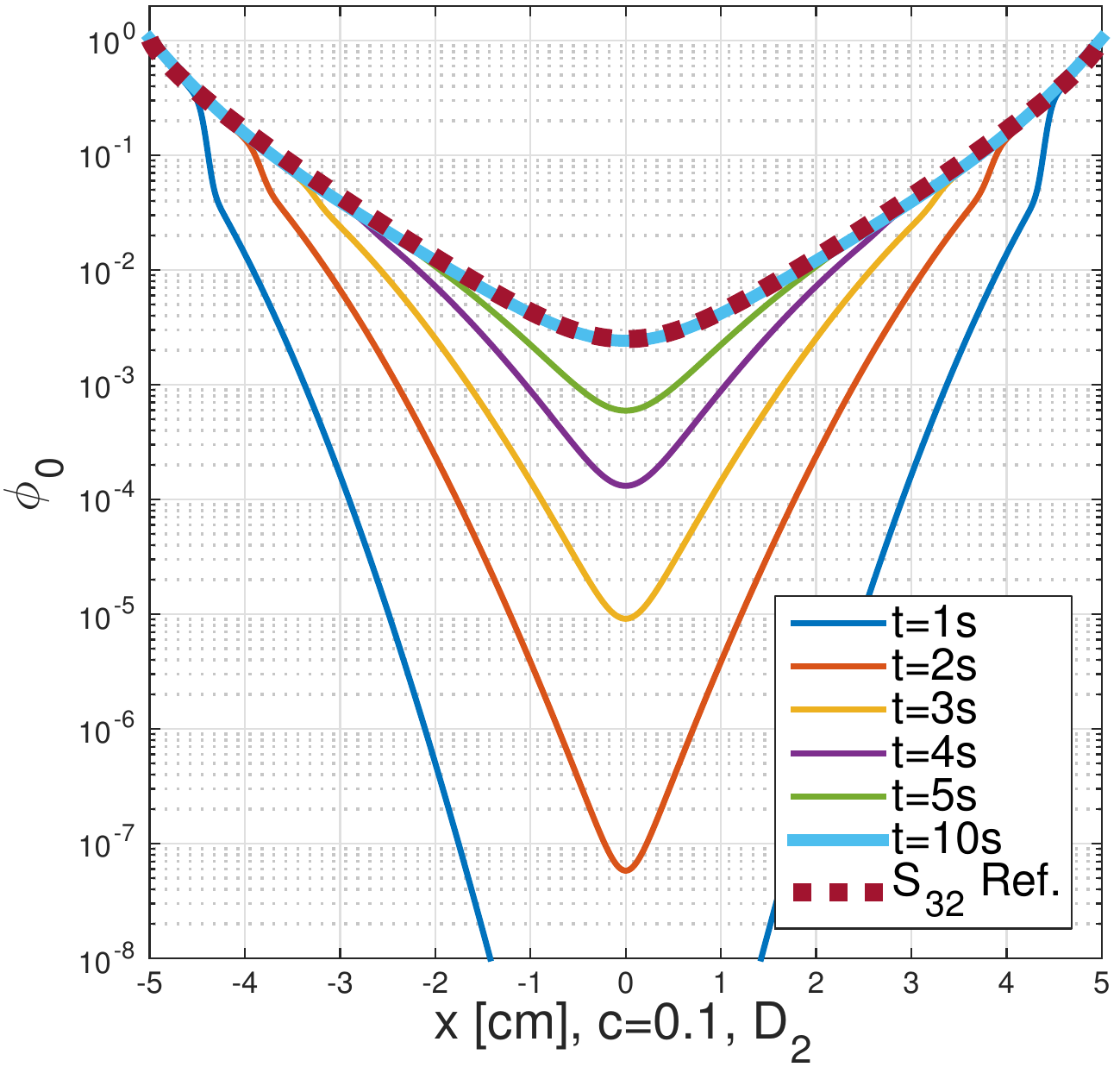}
	\end{subfigure}
	\caption{Comparison between TP$_2$ and D$_2$\ (P$_3$QS) solutions to two-beam problem.}
	\label{exp}
\end{figure}

The second test problem is a highly absorbing problem with isotropic incident angular fluxes on both sides of a slab. The scattering ratio, $c=\sigmas/\sigmat$,~of the medium is $0.1$. The original test problem is in steady state\cite{brunnerentropy}.\ We, however, run this problem in time-dependent mode to see how different methods approach the steady state solution. These results are shown in Figure\ \ref{exp}.

Both \tp{N}\ and \dn\ converge to the reference solution at 10\ s. Theoretically, incident particles from different sides of the slab are not supposed to meet before $t=L/(2v)=5$\ s. Yet, D$_2$\ artificially moves particles faster than their physical speeds, making the solution greater than $10^{-8}$ at $x=0$ as early as $t=2$s.  On the other hand, the TP$_2$\ model retains a sharper wavefront and as a result the solution at $x=0$ is below $10^{-8}$ until 5 s. 

Though the incident flux is isotropic on the boundary, the angular flux gradually turns to become strongly anisotropic and form a beam-like distribution in the middle of the slab due to the strong absorption. This beam-like behavior of the angular flux is a potential challenge for the model. Some closure models, such as the entropy based closure model (M$_N$),~have difficulty in resolving the beam. For the M$_N$~method, it tends to have artificial shock in the middle (Ref.~\cite{coryentropy,brunnerentropy}).~It is also suspected in Ref.~\cite{coryentropy}~that this shock could possibly caused by small errors when solving minimization problem governing the M$_N$~method. Fortunately, the TP$_{N}$~model does not have the artificial shock in this problem.

\subsection{Reed's problem}
The last test problem in this work is  Reed's problem \cite{reed_1971}. It contains several regions with largely varied properties including strong pure absorbers, voids, strong source and material discontinuities.

\begin{figure}[ht!]
	\begin{center}
		\includegraphics[width=.5\textwidth]{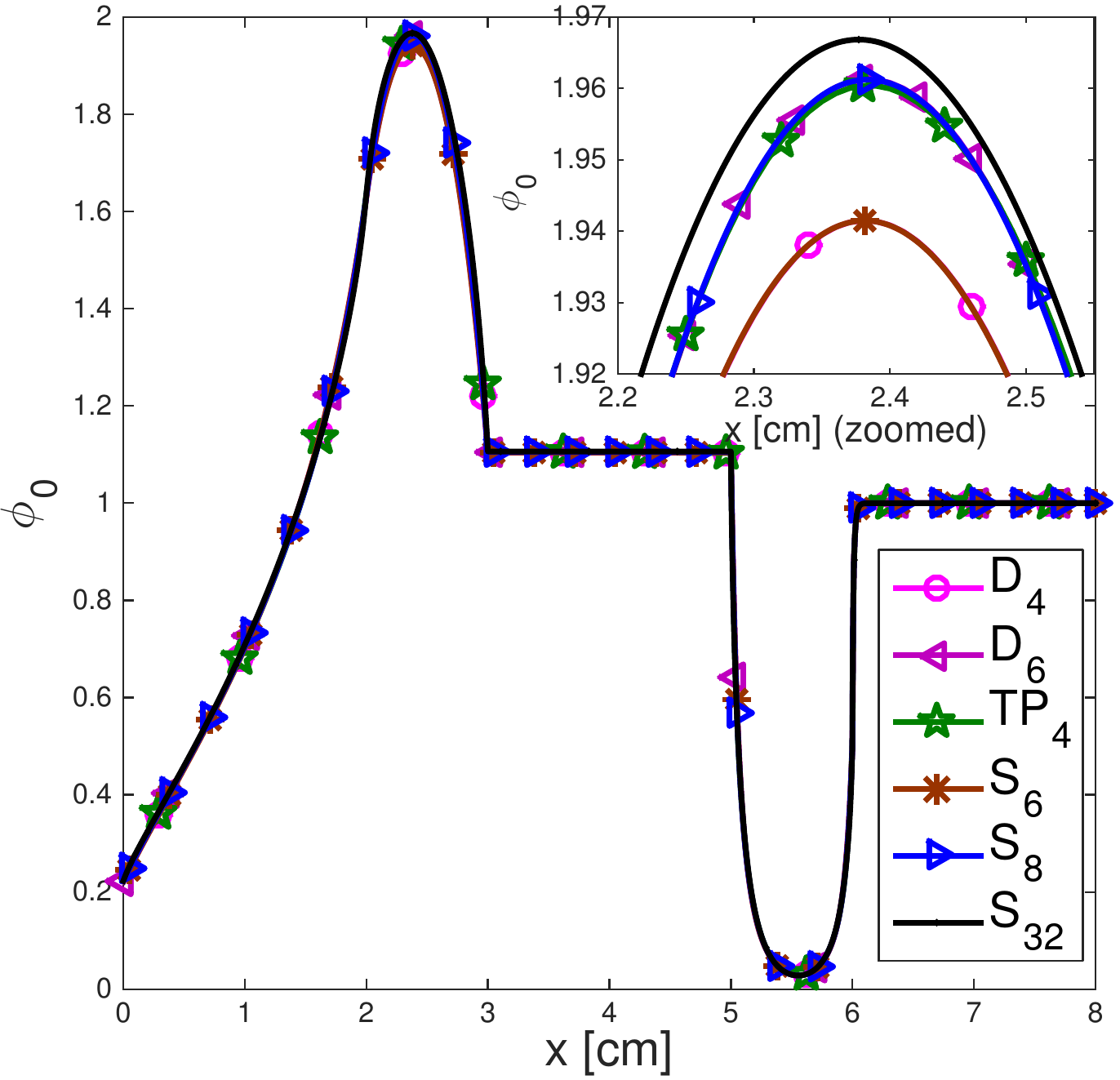}
		\caption[]{Reed's problem solved with D$_4$,\ TP$_4$,\ and D$_6$\ compared with S$_{32}$.}
		\label{reed2}
	\end{center}
\end{figure}

The numerical example in Figure~\ref{reed2}~is TP$_4$\ and D$_4$\ solutions of Reed's problem. In voids, in order to make the diffusive closure well-posed, an artificial absorption $\zeta$ is chosen:
\begin{equation}
	\st'=\st+\zeta, 
\end{equation}
where $\zeta$~is a small number, which is fixed at $10^{-8}$. For the TP$_N$\ model we only need a correction when $\tilde\sigma$ is zero, (i.e., in voids when then the scalar flux is constant in space and time). The correction we use is
\begin{equation}\label{tp}
\tilde{\sigma}'=\tilde{\sigma}+\zeta, 
\end{equation}

We use $800$\ cells in the discretization for D$_4$,\ D$_6$,\ and \tp{4}.\ 
The  S$_N$\ solutions are calculated with cell centered difference  using 16,000\ cells. We observe that TP$_4$\ retains an accuracy comparable to D$_6$\ and S$_8$. As a comparison, D$_4$\ displays comparable accuracy to S$_6$.\ Given that in 1-D slabs, S$_{N+1}$ gives identical solutions to P$_N$, this result is evidence that the \dn~and T\pn~models improve the solution as indicated by our residual analysis. As our analysis also predicts, the T\pn~solution is superior to both \dn~and \pn.

\begin{figure}[ht!]
	\begin{subfigure}{.5\textwidth}
	\begin{center}
		\hspace*{-1cm}\includegraphics[width=1.\textwidth]{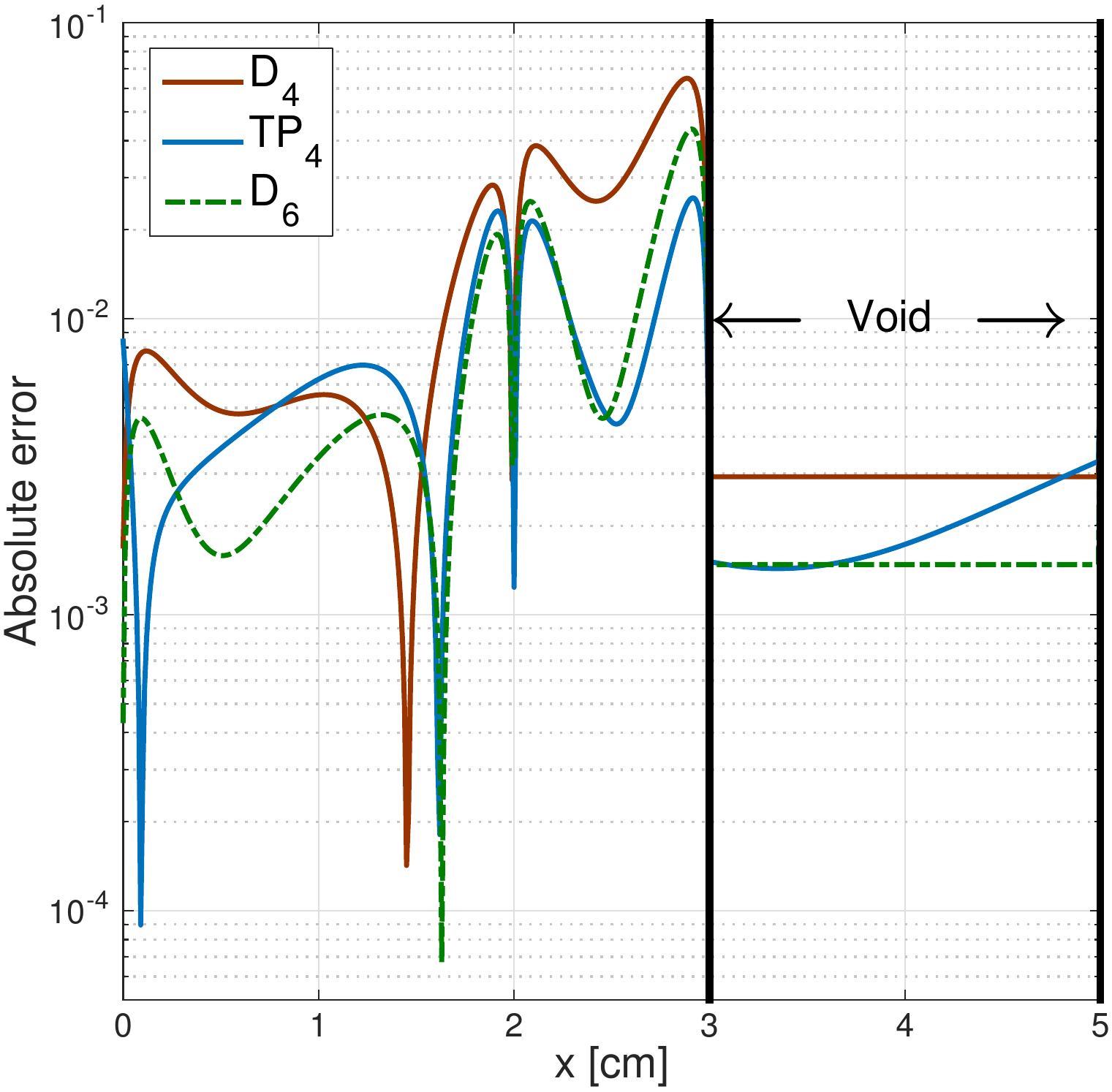}
		\caption[]{Pointwise absolute errors of D$_4$,\ TP$_4$,\ and D$_6$\ methods (Part 1).}
		\label{reed_er1}
	\end{center}
\end{subfigure}
~
\begin{subfigure}{.5\textwidth}
	\begin{center}
		\includegraphics[width=1.\textwidth]{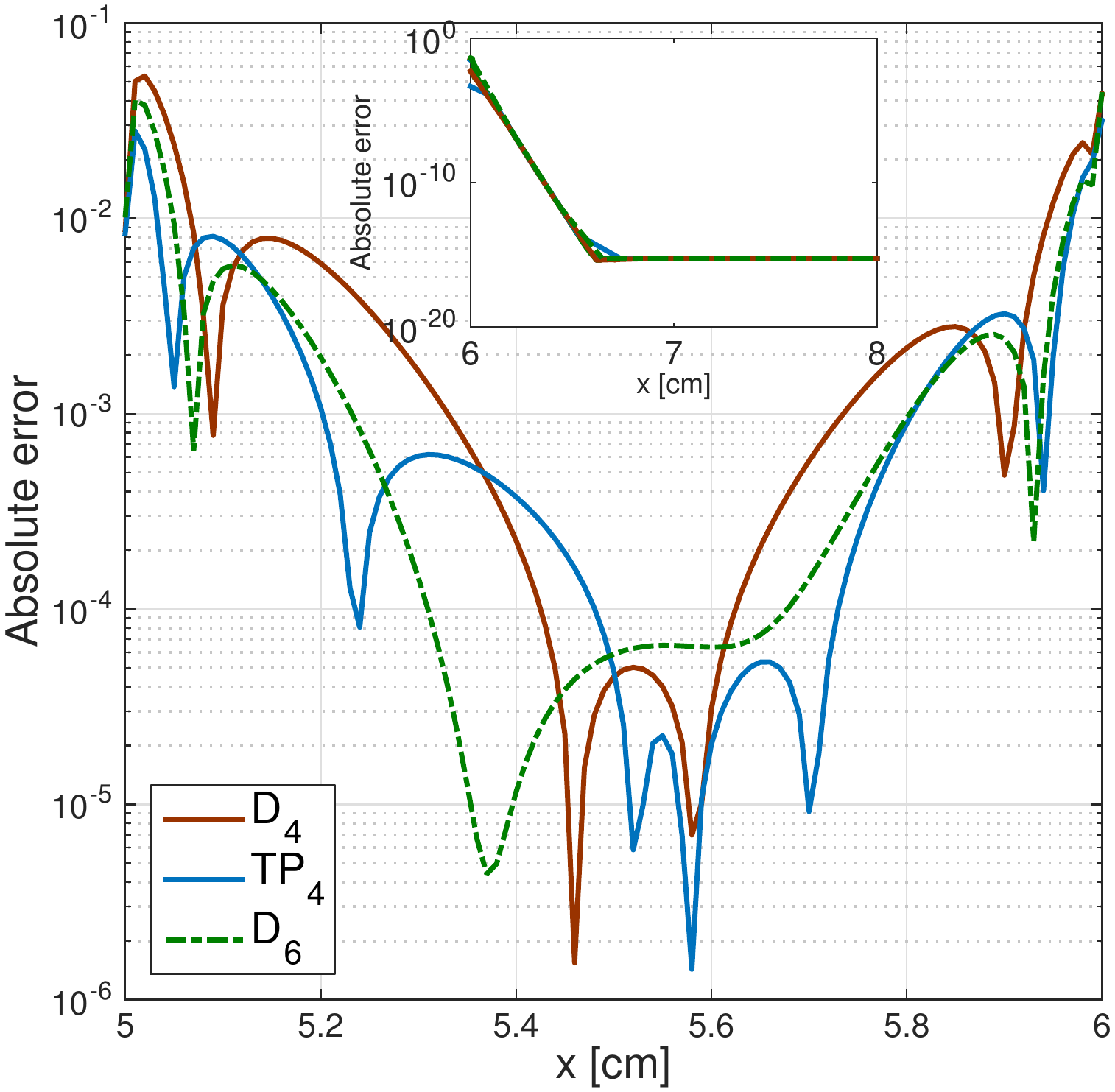}
		\caption[]{Pointwise absolute errors of D$_4$,\ TP$_4$,\ and D$_6$\ methods (Part 2).}
		\label{reed_er2}
	\end{center}
\end{subfigure}
\caption[]{\label{reed_error}Errors as a function of space in Reed's problem.}
\end{figure}

{Overall, as illustrated in Figure\ \ref{reed_error},\ the pointwise errors from TP$_4$\ are comparable to D$_6$\ and  smaller than those from D$_4$\ method in most regions especially for regions with large errors ($>10^{-2}$). We also observe that the boundary treatment in Eq~\eqref{e:bdy}\ brings about 0.8\% of error, slightly higher than D$_4$.\ However, the global $L_1$ norm of error for D$_4$ (estimated based on the fine-mesh S$_{32}$\ solution) is $0.129$\ and is larger than that of TP$_4$,\ which is $0.061$.\ For comparison, the D$_6$\ solution has an error of $0.066$.}

\subsection{2D line source problem}
\begin{figure}[ht!]
	\begin{subfigure}{.5\textwidth}
		\centering
		\hspace*{-1cm}\includegraphics[width=1.\linewidth]{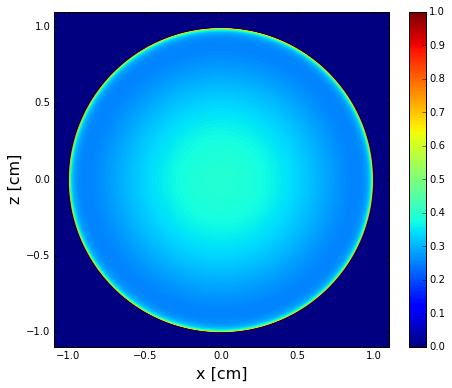}
		\caption{Analytic transport}
		\label{f:trans}
	\end{subfigure}
	~
	\begin{subfigure}{.5\textwidth}
		\centering
		\includegraphics[width=1.\linewidth]{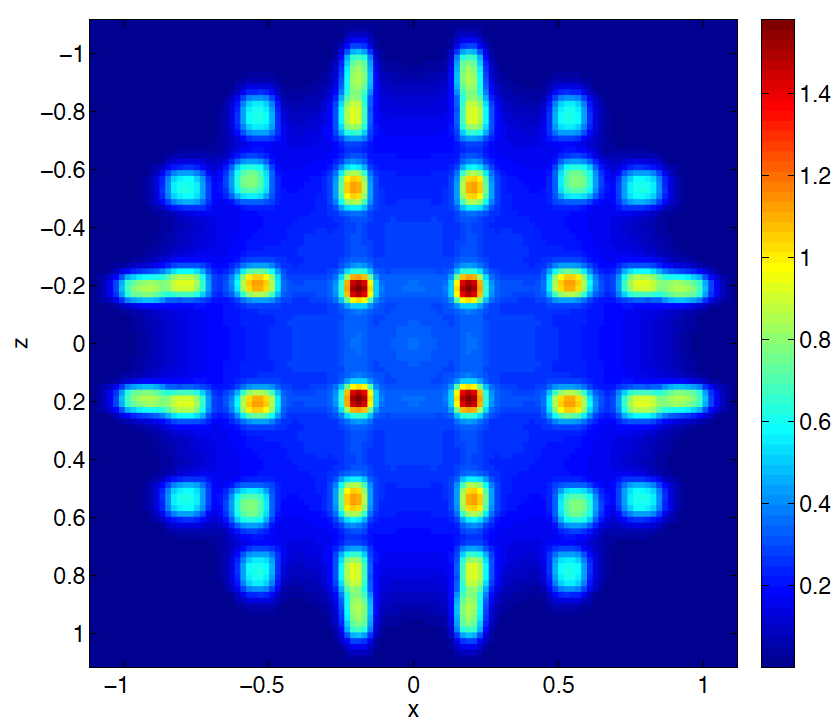}
		\caption{S$_8$}
		\label{f:s8}
	\end{subfigure}
	~
	\begin{subfigure}{.5\textwidth}
		\centering
		\hspace*{-1cm}\includegraphics[width=1.\linewidth]{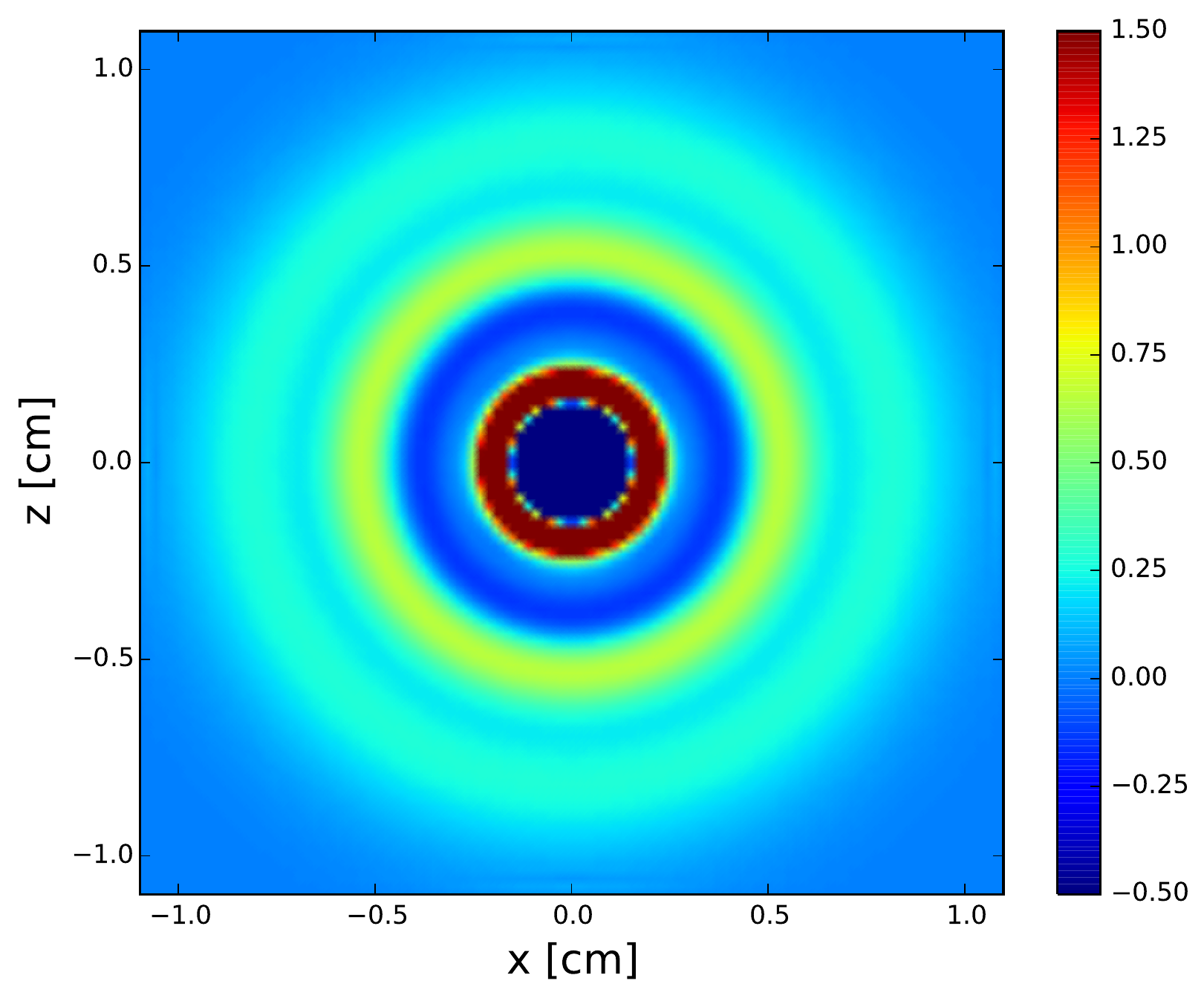}
		\caption{P$_7$}
		\label{f:p7}
	\end{subfigure}
	~
	\begin{subfigure}{.5\textwidth}
		\centering
		\includegraphics[width=1.\linewidth]{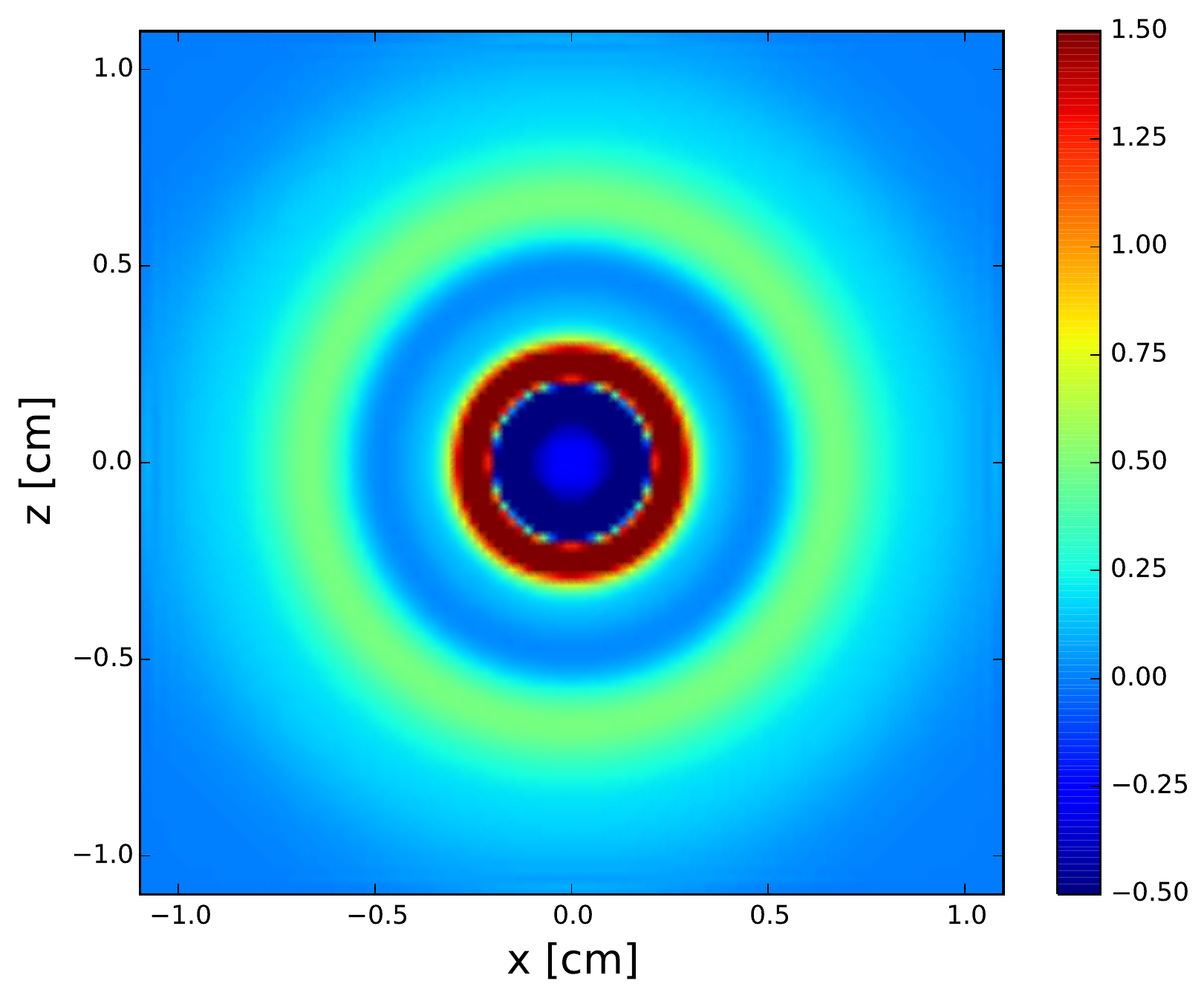}
		\caption{D$_6$}
		\label{f:d6}
	\end{subfigure}
	~
	\begin{subfigure}{.5\textwidth}
		\centering
		\hspace*{-1cm}\includegraphics[width=1.\linewidth]{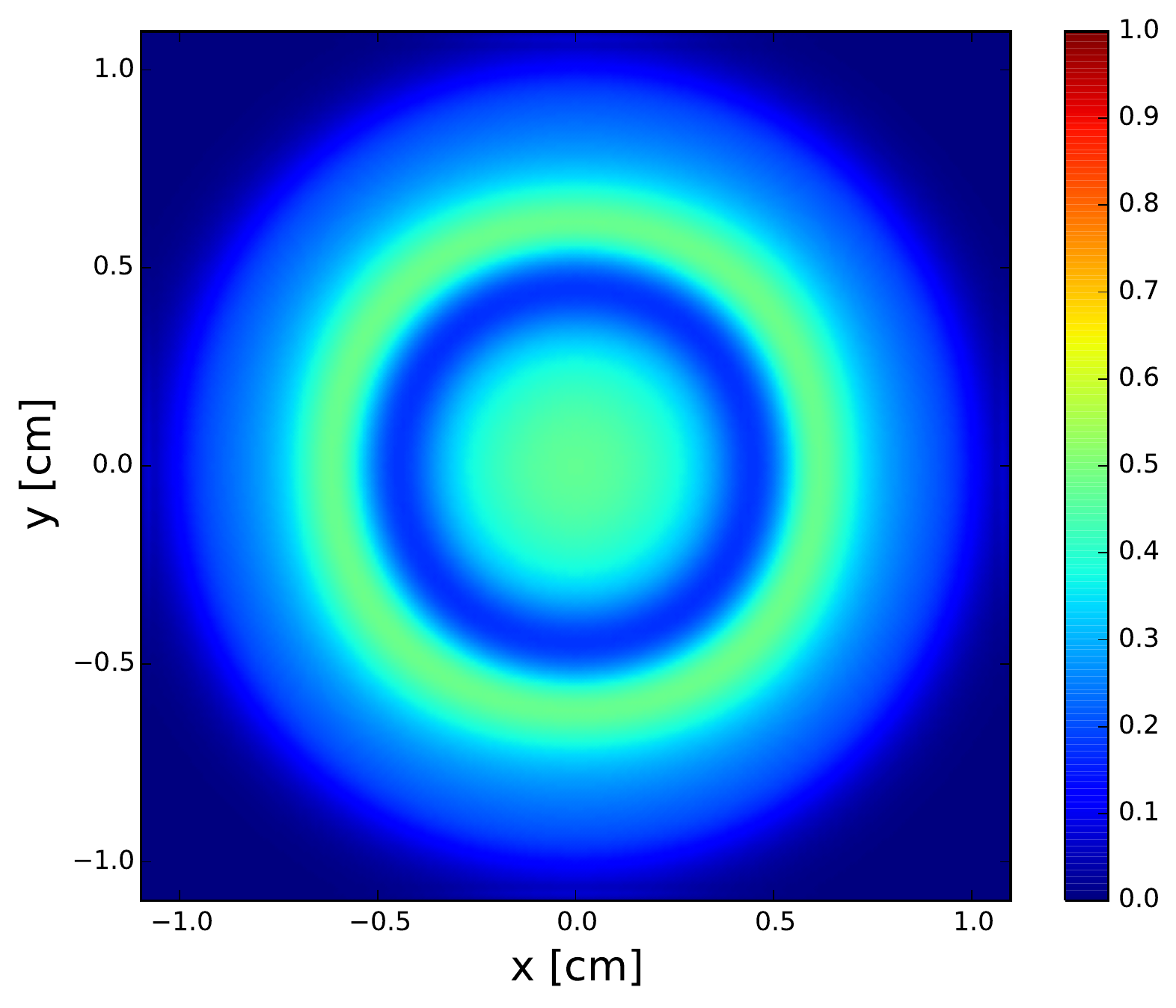}
		\caption{TP$_2$}
		\label{f:tp2}
	\end{subfigure}
	~
	\begin{subfigure}{.5\textwidth}
		\centering
		\includegraphics[width=1.\linewidth]{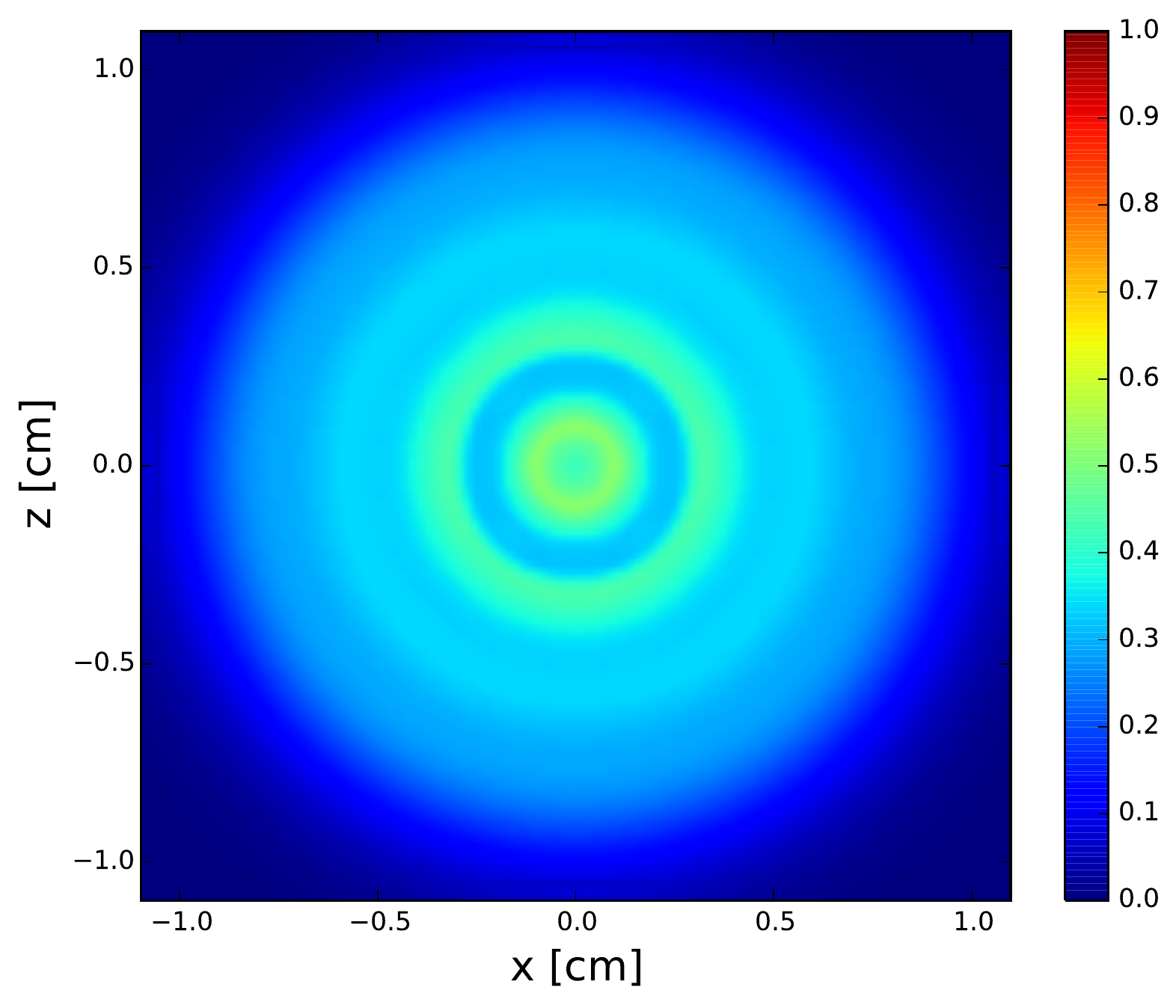}
		\caption{TP$_6$}
		\label{f:tp6}
	\end{subfigure}
	\caption{2D line source problem at $t=1$\ s.}
	\label{f:line}
\end{figure}
{
The line source problem is a 2D variation of the plane source problem in 1D slab geometry. The problem is an infinite, pure scattering medium ($\st=\sigmas=1$) with no source. The initial condition is given by\cite{ganapol}:
	\begin{equation}
	\psi(x,z,\hat{\Omega},0)=\frac{\delta(x)\delta(z)}{4\pi}
	\end{equation}
\pn,\ \dn\ and T\pn\ results in Figure\ \ref{f:line}\ are achieved with $\Delta t=0.02$\ s and $\Delta x=0.02$\ cm and S$_8$\ result is achieved from Ref.\ \cite{mccfpn09}.\ The analytic solution is shown in Figure\ \ref{f:trans}\ from the benchmark code AZURV1\cite{ganapol}.\ The wavefront at $r=\sqrt{x^2+z^2}=vt$,\ essentially a moving delta function, will induce oscillations and negative scalar fluxes in \pn\ and \dn\ methods as illustrated in Figures\ \ref{f:p7}\ and \ref{f:d6}.\ Meanwhile, for this streaming dominated problem, \sn\ results have strong ray-effects as in Figure\ \ref{f:s8}.\ On the other hand, TP$_2$\ solution presents plausible results in Figure\ \ref{f:tp2}.\ Increasing the angular order to TP$_6$\ will further improve the solution as illustrated in Figure\ \ref{f:tp6}. }

{Unlike the 1D T\pn\ model, multi-D T\pn\ models for different angular orders do not have a single coefficient $\alpha$,\ therefore, the results shown in Figure\ \ref{f:line}\ are obtained with different coefficients. Figure\ \ref{f:tp-lines}\ presents the diagonal lineout plots of TP$_2$,\ TP$_4$\ and TP$_6$\ with different $\alpha$.\ Changing the $\alpha$\ for each angular order can effectively change the results, as observed in 1D. Unfortunately, the ``optimal" values for different angular orders are different, e.g. while $\alpha=0.1$\ would lead to relatively accurate TP$_2$\ results, the ``optimal" $\alpha$\ changes to $1/3$\ and $1.5$\ for TP$_4$\ and TP$_6$,\ respectively. This should be addressed in future work.
\begin{figure}[ht!]
	\begin{subfigure}{.5\textwidth}
		\begin{center}
			\hspace*{-1cm}\includegraphics[width=1.\textwidth]{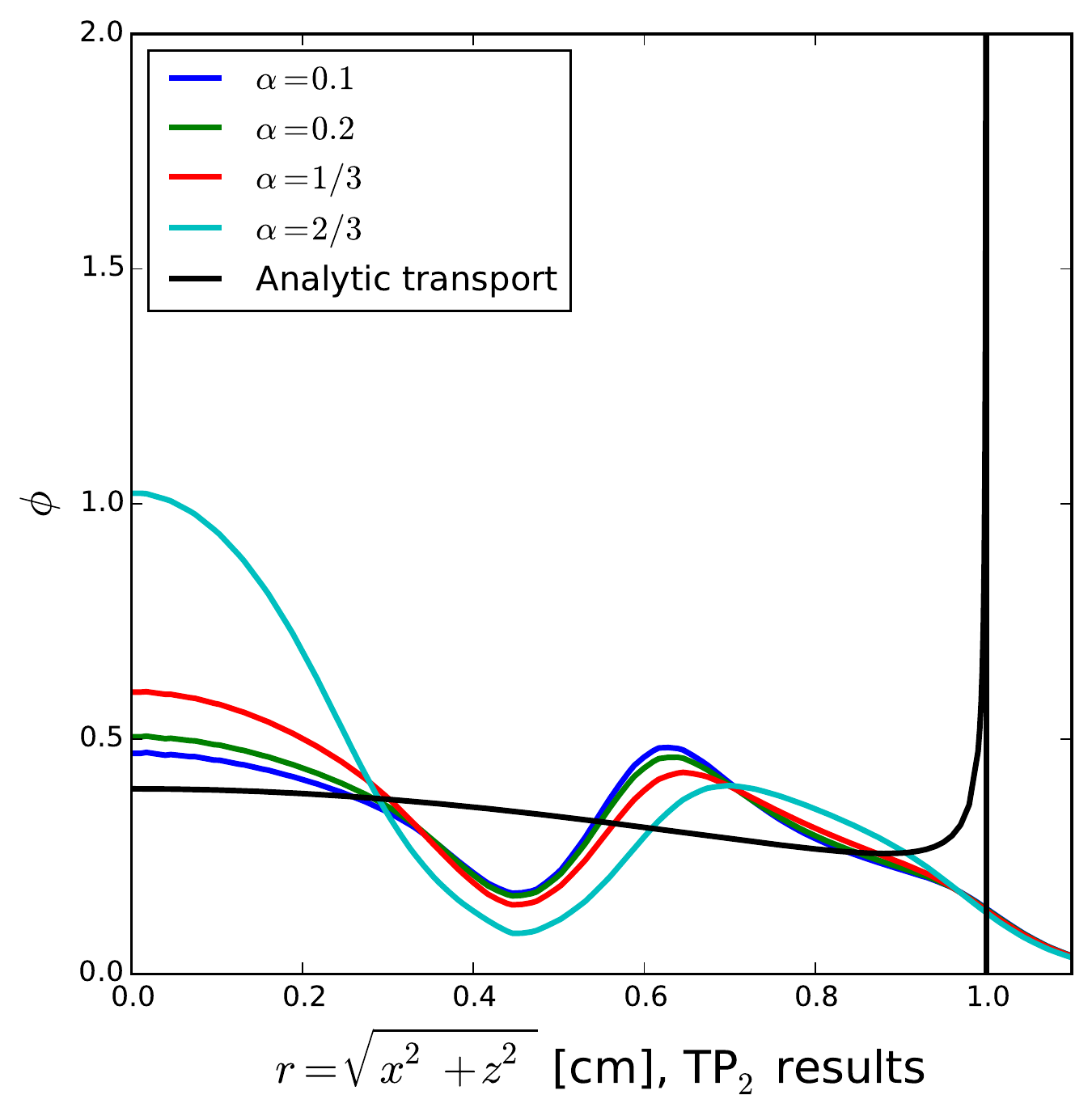}
			\caption[]{TP$_2$\ results with different $\alpha$.}
			\label{f:tp2-line}
		\end{center}
	\end{subfigure}
	~
	\begin{subfigure}{.5\textwidth}
		\begin{center}
			\hspace*{0cm}\includegraphics[width=1.\textwidth]{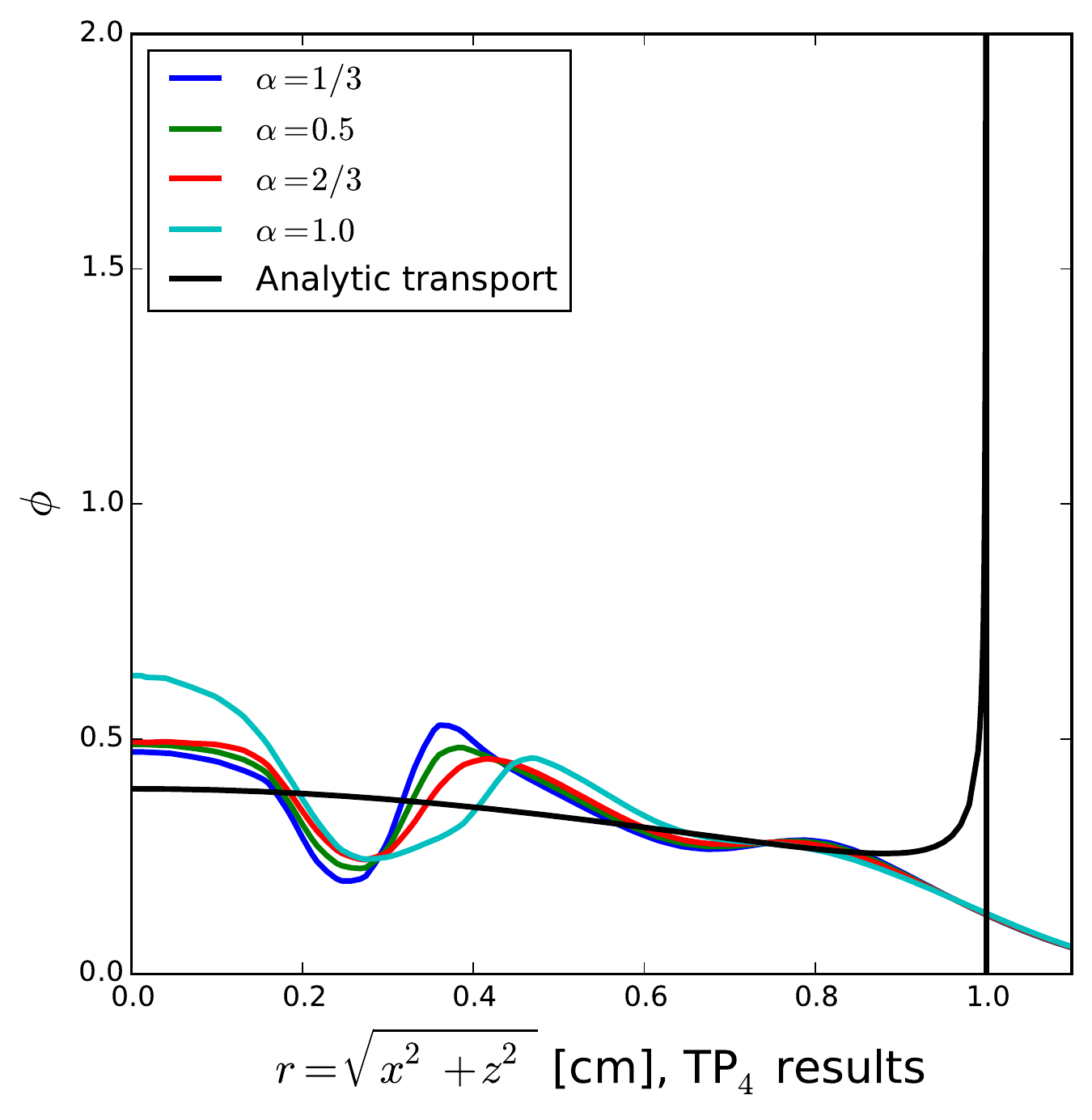}
			\caption[]{TP$_4$\ results with different $\alpha$.}
			\label{f:tp4-line}
		\end{center}
	\end{subfigure}
	~
	\begin{subfigure}{.5\textwidth}
		\begin{center}
			\hspace*{-1cm}\includegraphics[width=1.\textwidth]{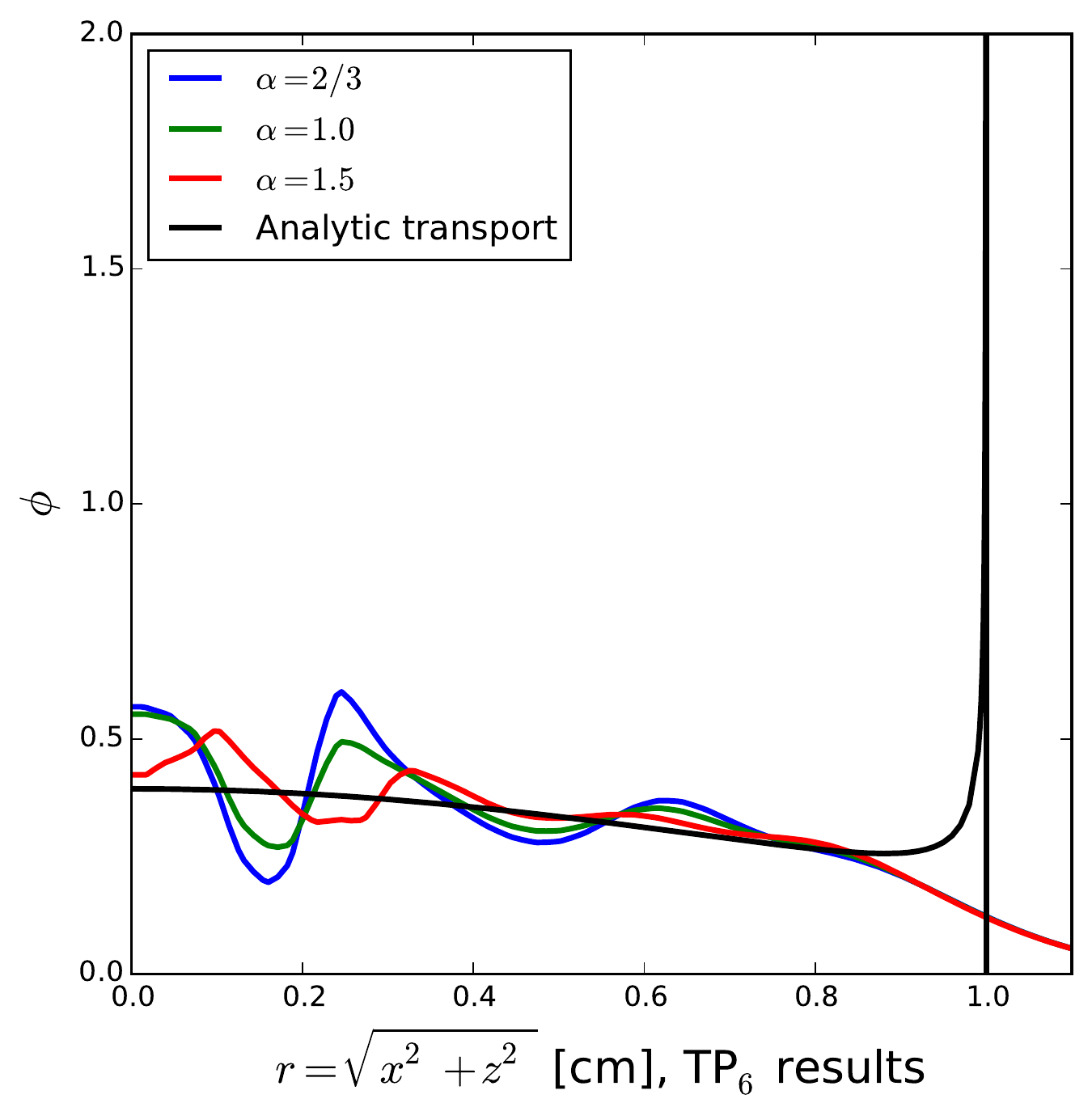}
			\caption[]{TP$_6$\ results with different $\alpha$.}
			\label{f:tp6-line}
		\end{center}
	\end{subfigure}
	\caption[]{\label{f:tp-lines}Diagonal lineout plots for TP$_2$\ and TP$_6$.}
\end{figure}
	}

{
2D \pn,\ \dn\ and T\pn\ are solved with (semi-)implicit discretization in time and DG/LDG method in space. A GMRES solver is used with the Jacobi preconditioner. A comparison of timings for T\pn\ and linear closures is made in Table\ \ref{t:cpu_time}.\ Simulations were run on a Mac mini with Intel i7-3615 processor and 16GB 1600MHz DDR3 RAM.
}

{
With LDG method, \dn\ and T\pn\ have the same number of degrees of freedom (DoFs) as P$_{N+1}$\ with the same basis functions used on the same mesh. The overall CPU time of T\pn\ is much shorter than \dn.\ The hypothesis is that the \dn\ model sets the time dependence of $\phi_{N+1}^m$\ to be zero, forcing particles in that mode to move with infinite speed. This makes \dn\ model physically ill-posed in time dependent problems. Numerically, the ill-posedness causes the degradation of the preconditioning efficiency.
}

{
On the other hand, though TP$_2$'s\ solving time is around 73\%\ higher than P$_3$,\ TP$_6$'s\ solving time is comparable to P$_7$. The correction brought by flux limiters affects not only the physical properties as discussed in 1D scenarios, but also the computational properties. 
Overall, we conclude T\pn\ would be comparably efficiently solved as \pn\ in multi-D applications.
\begin{table}[h]
	\centering
	\caption{Timings for line source problem at 1s.}
	\label{t:cpu_time}
	\begin{tabular}{|c|c|c|c|}
		\hline
		& P$_3$ & TP$_{2}$ & D$_{2}$\\
		\hline
Setup+assembly time [s] & 6.9 & 7.4 & 7.2\\
\hline
		Estimate+assemble limiter time [s]& 0 & 16.5 & 0 \\
		\hline
		Solving+preconditioning time [s]& 124.6 & 215.6 & 562.8 \\
		\hline
		Total CPU time [s]& 131.5  & 228.5 & 570.0\\
		\hline
		\hline
		& P$_5$ & TP$_{4}$ & D$_{4}$\\
		\hline
Setup+assembly time [s] & 20.6  & 19.5 & 18.5\\
\hline
Estimate+assemble limiter time [s]& 0 & 28.3 & 0 \\
\hline
Solving+preconditioning time [s]& 383.5 & 482.0 &  1308.8\\
\hline
	Total CPU time [s]& 404.1 & 529.8 & 1327.3\\
		\hline
		\hline
		& P$_7$ & TP$_{6}$ & D$_{6}$\\
		\hline
		Setup+assembly time [s] & 43.7  & 42.4 & 43.0\\
		\hline
		Estimate+assemble limiter time [s]& 0 & 153.2 & 0 \\
		\hline
		Solving+preconditioning time [s]& 734.5 & 760.4 & 2424.7 \\
		\hline
		Total CPU time [s]& 778.1 & 956.0 & 2467.7\\
		\hline
	\end{tabular}
\end{table}
	}

\section{Concluding remarks}
In this paper, we analyzed the effects on the \pn~approximation residual caused by different closures. We provide a new explanation of the reasons the conventional \pn~and \pn~with diffusive closure has issues in transient simulations, such as the pulsed plane source problem {in 1D and line source problem in 2D}. Based on the analysis, we proposed two novel closures, the ``moment-limited" closure for 1D and ``transient'' \pn\ closure in 1D and 2D. The results we presented indicate that, relative to other linear closures, our new closures perform better on a variety of problems, including the notorious plane source problem {and line source problem}. 

{
}

{
}

Finally, we believe that the ideas behind the T\pn~model could be used in other transport models. For instance, this type of closure could be applied to the simplified \pn~method \cite{McClarren:2011ga}. 

\section*{Acknowledgements}
The authors would like to thank the anonymous reviewers for their valuable comments and suggestions to improve the quality of the paper. W. Zheng would like to thank Dr.\ Wolfgang Bangerth from the Dept. of Mathematics at Texas A\&M University for helpful discussions about efficient implementations of the multi-D applications in deal.II. Also, W. Zheng is thankful to Dr.\ Robert Lowrie from Los Alamos National Laboratory for providing the LDG idea for multi-D application.

This project is funded by Department of Energy NEUP research grant from Battelle Energy Alliance, LLC- Idaho National Laboratory, Contract No: C12-00281.

\section*{References}
\bibliography{mybibfile}

\begin{thebibliography}{10}
\expandafter\ifx\csname url\endcsname\relax
  \def\url#1{\texttt{#1}}\fi
\expandafter\ifx\csname urlprefix\endcsname\relax\def\urlprefix{URL }\fi
\expandafter\ifx\csname href\endcsname\relax
  \def\href#1#2{#2} \def\path#1{#1}\fi

\bibitem{glasstone}
G.~I. Bell, S.~Glasstone, Nuclear Reactor Theory, 3rd Edition, Krieger Pub Co,
  Princeton, NJ, 1985.

\bibitem{pomraning1973equations}
G.~Pomraning, \href{https://books.google.com/books?id=EFbpgt-Rlh8C}{The
  Equations of Radiation Hydrodynamics}, Dover books on physics, Dover
  Publications, 1973.
\newline\urlprefix\url{https://books.google.com/books?id=EFbpgt-Rlh8C}

\bibitem{Grad:1949wi}
H.~Grad, {On the kinetic theory of rarefied gases}, Communications on Pure and
  Applied Mathematics 2~(4) (1949) 331--407.

\bibitem{markowich2012semiconductor}
P.~A. Markowich, C.~A. Ringhofer, C.~Schmeiser,
  \href{https://books.google.com/books?id=nCr7CAAAQBAJ}{Semiconductor
  Equations}, Springer Vienna, 2012.
\newline\urlprefix\url{https://books.google.com/books?id=nCr7CAAAQBAJ}

\bibitem{Mathews:1999uv}
K.~A. Mathews, {On the propagation of rays in discrete ordinates}, Nuclear
  science and engineering 132~(2) (1999) 155--180.

\bibitem{Morel:2003vt}
J.~E. Morel, T.~A. Wareing, R.~B. Lowrie, D.~K. Parsons, {Analysis of
  ray-effect mitigation techniques}, Nuclear science and engineering 144~(1).

\bibitem{brunner_app_rad_trans}
T.~A. Brunner, {Forms of Approximate Radiation Transport}, Tech. Rep.
  SAND2002-1778, Sandia National Laboratory (2002).

\bibitem{mccfpn09}
R.~G. McClarren, C.~D. Hauck,
  \href{http://www.sciencedirect.com/science/article/pii/S0375960110002021}{Simulating
  radiative transfer with filtered spherical harmonics}, Physics Letters A
  374~(22) (2010) 2290 -- 2296.
\newblock \href {http://dx.doi.org/http://dx.doi.oR.
  G./10.1016/j.physleta.2010.02.041} {\path{doi:http://dx.doi.oR.
  G./10.1016/j.physleta.2010.02.041}}.
\newline\urlprefix\url{http://www.sciencedirect.com/science/article/pii/S0375960110002021}

\bibitem{McClarren:2008cq}
R.~G. McClarren, T.~M. Evans, R.~B. Lowrie, J.~D. Densmore, {Semi-implicit time
  integration for thermal radiative transfer}, Journal of Computational Physics
  227~(16) (2008) 7561--7586.

\bibitem{coryentropy}
C.~D. Hauck, High-order entropy-based closures for linear transport in slab
  geometry, Commun. Math. Sci 9~(1) (2011) 187--205.

\bibitem{McClarren:2008hu}
R.~G. McClarren, J.~P. Holloway, T.~A. Brunner, {On solutions to the P$_n$
  equations for thermal radiative transfer}, Journal of Computational Physics
  227~(5) (2008) 2864--2885.

\bibitem{mccdissertation}
R.~G. McClarren, {Spherical Harmonics Method for Thermal Radiation Transport},
  Ph.D. dissertation, University of Michigan (2007).

\bibitem{levermoredn}
M.~Sch\"{a}fer, M.~Frank, C.~D. Levermore, {Diffusive Corrections to PN
  Approximations}, Multiscale Modeling \& Simulation 9~(1) (2011) 1--28.
\newblock \href {http://dx.doi.org/10.1137/090764542}
  {\path{doi:10.1137/090764542}}.

\bibitem{p3qs}
K.~S. Oh, J.~P. Holloway,
  \href{http://www.scopus.com/inward/record.url?eid=2-s2.0-74549196287&partnerID=40&md5=067624efc28e6ab8fcd497d7c581617a}{{A
  Quasi-static Closure for 3rd Order Spherical Harmonics Time-Dependent
  Radiation Transport in 2-D}}, Vol.~3, 2009, pp. 1938--1948, cited By 0.
\newline\urlprefix\url{http://www.scopus.com/inward/record.url?eid=2-s2.0-74549196287&partnerID=40&md5=067624efc28e6ab8fcd497d7c581617a}

\bibitem{Olson:2000vq}
G.~L. Olson, L.~H. Auer, M.~L. Hall, {Diffusion, P$_1$, and other approximate
  forms of radiation transport}, Journal of Quantitative Spectroscopy and
  Radiative Transfer 64~(6) (2000) 619--634.

\bibitem{cory_hauck_closures}
C.~K. Garrett, C.~D. Hauck, {A Comparison of Moment Closures for Linear Kinetic
  Transport Equations: The Line Source Benchmark}, Transport Theory and
  Statistical Physics 42~(6-7) (2013) 203--235.
\newblock \href {http://dx.doi.org/10.1080/00411450.2014.910226}
  {\path{doi:10.1080/00411450.2014.910226}}.

\bibitem{brunnerentropy}
T.~A. Brunner, J.~P. Holloway,
  \href{http://www.sciencedirect.com/science/article/pii/S0022407300000996}{One-dimensional
  {Riemann} solvers and the maximum entropy closure}, Journal of Quantitative
  Spectroscopy and Radiative Transfer 69~(5) (2001) 543 -- 566.
\newblock \href {http://dx.doi.org/http://dx.doi.oR.
  G./10.1016/S0022-4073(00)00099-6} {\path{doi:http://dx.doi.oR.
  G./10.1016/S0022-4073(00)00099-6}}.
\newline\urlprefix\url{http://www.sciencedirect.com/science/article/pii/S0022407300000996}

\bibitem{ppn}
C.~D. Hauck, R.~G. McClarren, {Positive P$_N$ Closures}, SIAM Journal on
  Scientific Computing 32~(5) (2010) 2603--2626.
\newblock \href {http://dx.doi.org/10.1137/090764918}
  {\path{doi:10.1137/090764918}}.

\bibitem{McClarren:2010de}
R.~G. McClarren, C.~D. Hauck, {Robust and accurate filtered spherical harmonics
  expansions for radiative transfer}, Journal of Computational Physics 229~(16)
  (2010) 5597--5614.

\bibitem{fpn_radice}
D.~Radice, E.~Abdikamalov, L.~Rezzolla, C.~D. Ott, A new spherical harmonics
  scheme for multi-dimensional radiation transport i. static matter
  configurations, Journal of Computational Physics 242 (2013) 648 -- 669.
\newblock \href {http://dx.doi.org/http://dx.doi.oR.
  G./10.1016/j.jcp.2013.01.048} {\path{doi:http://dx.doi.oR.
  G./10.1016/j.jcp.2013.01.048}}.

\bibitem{ahrens_fpn}
C.~Ahrens, S.~Merton, {An Improved Filtered Spherical Harmonic Method for
  Transport Calculations}, 2013.

\bibitem{reed_1971}
W.~Reed, New difference schemes for the neutron transport equation, Nucl. Sci.
  Eng. 46: No. 2, 309-14.

\bibitem{Morel:2000vh}
J.~E. Morel, {Diffusion-limit asymptotics of the transport equation, the
  P$_{1/3}$ equations, and two flux-limited diffusion theories}, Journal of
  Quantitative Spectroscopy and Radiative Transfer 65~(5) (2000) 769--778.

\bibitem{taylor_ans12}
R.~G. McClarren, T.~K. Lane, {A Flux-Limited Diffusion Method for Simulating
  Radiative Shocks}, in: {ANS Winter Meeting 2012}, Vol. 107, ANS, 2012, pp.
  557--559, san Diego, CA, November.

\bibitem{cockburn_ldg}
B.~Cockburn, C.-W. Shu, {The Local Discontinuous Galerkin finite element method
  for convection–-diffusion systems}, SIAM J. Numer. Anal. 35 (1998)
  2440--2463.
\newblock \href {http://dx.doi.org/10.1137/090764918}
  {\path{doi:10.1137/090764918}}.

\bibitem{dealii82}
W.~Bangerth, T.~Heister, L.~Heltai, G.~Kanschat, M.~Kronbichler, M.~Maier,
  B.~Turcksin, T.~D. Young, The \texttt{deal.II} library, version 8.2, Archive
  of Numerical Software 3.

\bibitem{ganapol}
B.~D. Ganapol, R.~S. Baker, J.~A. Dahl, R.~E. Alcouffe, Homogeneous infinite
  media time-dependent analytical benchmarks, Tech. Rep. LA-UR--01-1854, Los
  Alamos National Laboratory (2001).

\bibitem{McClarren:2011ga}
R.~G. McClarren, {Theoretical Aspects of the Simplified P$_n$ Equations},
  Transport Theory and Statistical Physics 39~(2-4) (2011) 73--109.

\end{thebibliography}

\end{document}